\documentstyle[aps]{revtex}
\begin{document}
\draft
\title{Lattice vibrations of $\alpha'$-NaV$_2$O$_5$ in the low-temperature
phase. Magnetic bound states?} \author{M. N. Popova, A. B. Sushkov, S. A.
Klimin, E. P. Chukalina} \address{Institute of Spectroscopy of
Russian Academy of Sciences, \\ 142190 Troitsk, Moscow reg., Russia}
\author{B. Z. Malkin}
\address{Kazan State University, 420008 Kazan, Russia}
\author{M. Isobe, Yu. Ueda}
\address{Institute for Solid State Physics, The University of Tokyo \\
7--22--1 Roppongi, Minato-ku, Tokyo 106, Japan}
\date{\today}
\maketitle

\begin{abstract}
We report high resolution polarized infrared studies of the quarter-filled
spin ladder compound $\alpha'$-NaV$_2$O$_5$ as a function of temperature
(5~K$\le T\le300$~K). Numerous new modes were detected below the
temperature $T_c=34$~K of the phase transition into a charge ordered
nonmagnetic state accompanied by a lattice dimerization.
We analyse the Brillouin zone (BZ) folding due to lattice
dimerization at $T_c$ and show that some peculiarities of the
low-temperature vibrational spectrum come from quadruplets folded
from the BZ point ($\frac12, \frac12, \frac14$). We discuss an
earlier interpretation of the 70, 107, and 133~cm$^{-1}$ modes as
magnetic bound states and propose the alternative interpretation as
folded phonon modes strongly interacting with charge and spin
excitations.
\end{abstract}



\section{Introduction}

The magnetic insulator $\alpha'$-NaV$_2$O$_5$ has received a
considerable attention since 1996, when Isobe and Ueda  reported a
spin-Peierls-like transition at $T_c=35$~K \cite{c1}. The structure of this
vanadate consists of two-leg ladders formed by corner-sharing distorted
$V_2$O$_5$ pyramids running along the orthorhombic b-axis, with their
rungs along the a-axis (see Fig.~1). Neighboring ladders are linked via
common edges of the pyramids to form the $ab$-layers. Na atoms lie
between the layers. At room temperature, the structure is centrosymmetric
(space group $Pmmn$--$D_{2h}^{13}$) with only one crystallographic
position for vanadium having a formal valence
+4.5~\cite{smol,meet,vonschner,nav3,ludecke,ohama}. As a result, one
electron is shared by two V atoms and is actually distributed over a
V--O--V molecular orbital at the ladder rung. A quarter-filled ladder
along the $b$-direction is formed, which explains one-dimensional magnetic
properties of $\alpha'$-NaV$_2$O$_5$ above $T_c$~\cite{smol,horsch}.

Below $T_c$, a steep isotropic decrease of the magnetic susceptibility
corresponding to a singlet-triplet gap of $\Delta$=85~K~\cite{c1,weiden}
and a lattice dimerization described by the propagation wave vector ${\bf
k}=\left(\frac12 \frac12 \frac14\right)$~\cite{fujii} were recently found to
be accompanied by a charge ordering of a zigzag
type~\cite{ohama,nav5,luthi,x-anom}
The nature of the state below $T_c$ is presently under intense
investigation.  The recent synchrotron X-ray diffraction
studies of the low-temperature (LT)
structure~\cite{ludecke,deboer,thalm} resulted in the acentric space
group $Fmm2$--$C_{2v}^{18}$ with three essentially different positions for
the vanadium atoms arranged into a sequence of zigzag modulated ladders
separated by nonmodulated ones (see Fig.~1). The corresponding three
vanadium valence states, +4, +5, and +4.5, have been suggested from the
bond-valence analysis in Ref.~\cite{europhys}. This finding
contradicts the NMR observation of only two inequivalent electronic states
of vanadium below $T_c$~\cite{ohama}. More recent bond-valence calculations
suggest, however, only two vanadium valences despite the three
crystallographically inequivalent V sites~\cite{thalm}. Nevertheless,
several observations, in particular, the properties of X-ray anomalous
scattering~\cite{x-anom}, can be better explained if one assumes the
modulation of all the vanadium ladders, and therefore different
subgroups of $Fmm2$
were considered~\cite{ohama2,palstra}. In their Raman scattering study of
folded modes, the authors of Ref.~\cite{konst} claimed the centrosymmetric
$C2/c$ group for the LT-phase of $\alpha'$-NaV$_2$O$_5$, from purely
spectroscopic reasons.

Folded modes of the dimerized lattice of $\alpha'$-NaV$_2$O$_5$ have been
observed first in the far infrared (FIR) transmission~\cite{nav1},
then Raman scattering~\cite{weiden,kuroe,fischer} and FIR
reflection~\cite{dsmirnov1,dam2} measurements. Only two out of more
than thirty observed~\cite{nav1,mism,dsmirnov1,dam2} new
low-temperature FIR modes, namely, the modes at 718~cm$^{-1}$ (${\bf
E} \parallel {\bf a}, {\bf b}$)~\cite{dsmirnov1,dsmirnov2,dam2} and
960~cm$^{-1}$ (${\bf E}\parallel {\bf a}$)~\cite{dam2} have been
studied in detail. Three lowest frequency
Raman lines (67, 107, 133 cm$^{-1}$) that emerge below $T_c$ have been
attributed to magnetic singlet bound
states~\cite{fischer,lemmens,fischer2}. The recent work~\cite{konst3}
called this interpretation into question, however.
Thus, both the low-temperature structure of $\alpha'$-NaV$_2$O$_5$ and the
nature of the low-energy excitations still remain controversial. The
necessary steps towards solving these problems include (i) an acquisition
of precise high resolution experimental data on the properties of new
low-temperature infrared and Raman modes in all the principal
polarizations and (ii) a theoretical analysis of the properties of
modes folded into the Brillouin zone (BZ) center of the LT-phase from
different BZ-points of the paraphase.

The present work is devoted to the detailed high-resolution FIR
transmission study of the frequencies, widths, and shapes of additional
low-temperature modes of $\alpha'$-NaV$_2$O$_5$ for $T \le T_c$ in ${\bf
E}\parallel {\bf a}$, ${\bf E}\parallel{\bf b}$, and ${\bf E}\parallel{\bf
c}$ polarizations. To understand, at least qualitatively,
polarization properties, splittings, and temperature behavior of the
folded modes, we perform lattice dynamical calculations of the phonon
dispersion curves throughout the Brillouin zone of the paraphase and
of frequency changes due to lattice distortion and charge ordering
below $T_c$. Possible manifestation of magnetic excitations and of
the interaction between lattice, charge, and spin excitations
in the FIR spectra are discussed.

\section{Experiment}

Samples of stoichiometric $\alpha'$-NaV$_2$O$_5$ used in this study were
grown by a melt growth method using NaVO$_3$ as a flux~\cite{isobe}. The
process resulted in single crystals with dimensions from
$1\times5\times0.5$\,mm to $3\times17.3\times1.6$\,mm along the $a$-, $b$-,
and $c$-axes, respectively. Crystals from different batches were used. For
transmission measurements in the ${\bf k}\parallel{\bf c}$, ${\bf
E}\parallel {\bf a}$ or ${\bf E}\parallel{\bf b}$ configurations, we
prepared four thin samples cleaved perpendicular to the $c$-axis. Their
thicknesses were $111$, $45$, $14$, and $6$\,$\mu$m. For the measurements
in the ${\bf k}\parallel {\bf a}$, ${\bf E} \parallel{\bf c}$ or ${\bf
E}\parallel{\bf b}$ geometries, two samples were used. One of them was
1.3\,mm thick along the $a$-axes (sample 2). The other one was covered with
epoxy and then polished to the thickness of 150\,$\mu$m (sample 1). The
samples were checked by X-ray diffraction, magnetization, and ESR
measurements. They exhibited a sharp transition at about 34\,K.

The transmittance spectra in a polarized light incident normally to
the sample surface were measured in the spectral range
25--2000~cm$^{-1}$ at a resolution  0.05--1.0~cm$^{-1}$ using a BOMEM
DA3.002 Fourier transform spectrometer equipped with a helium-vapor optical
cryostat. Samples were mounted to a specially constructed cryostat inset
that partially compensated thermal variations of the cryostat length. The
temperature of the sample varied from 300 to 6\,K and was stabilized with
the precision of $\pm0.1$\,K. At each temperature, a reference spectrum was
taken. For the case of sample~1 measured in ${\bf E}\parallel{\bf c}$
polarization, transmittance is given relative to the layer of epoxy
having the same thickness as the sample. Precision of the
experimental line position was 0.05--0.5~cm$^{-1}$, depending on a
particular line. Precise absolute wave number scale is an intrinsic
property of Fourier transform spectroscopy.

\section{Results}

Fig.~\ref{fig2} shows FIR polarized transmittance in the spectral
range from 55 to 350~cm$^{-1}$ for
$\alpha'$-NaV$_2$O$_5$ at temperatures above and below $T_c$. At $T\simeq
40$~K$>T_c$ in the displayed frequency range we found two phonon lines
($\omega_{TO}=178$ and 225~cm$^{-1}$) in ${\bf
E}\parallel {\bf b}$ polarization (for the discussion of the
215~cm$^{-1}$ line see Ref.~\cite{nav3}); three strongly asymmetric
phonon peaks at 90, 140, and 254~cm$^{-1}$ visible on
a strong absorption background in ${\bf E}\parallel {\bf a}$ polarization;
two strong phonon lines with $\omega_{TO}=163$ and 180~cm$^{-1}$,
$\omega_{LO}=169$ and 215~cm$^{-1}$ (compare with $\omega_{TO}=162$ and
179~cm$^{-1}$, $\omega_{LO}=165$ and 212~cm$^{-1}$ as determined from FIR
reflectance spectra at room temperature~ \cite{nav3}) and an absorption
peak at 282~cm$^{-1}$ in ${\bf E} \parallel {\bf c}$ polarization.
This peak can be assigned to a missing $B_{1u}$ phonon placed at
298~cm$^{-1}$ by lattice dynamics calculations for the
high-temperature (HT)-phase~\cite{nav3} and not observed in FIR reflectance
spectra~\cite{nav3,dsmirnov2}.

A broad absorption band centered at about 320~$^{-1}$ is present in
${\bf E}\parallel {\bf a}$ spectrum (see also~\cite{nav1,nav2,dam}).
One can clearly see it in Fig.~2a where the transmittance of a
14~$\mu$m thick sample is almost zero near this frequency. No
broad-band absorption of a comparable intensity exist it
${\bf E }\parallel {\bf b}$ (Fig.~2b) and ${\bf E}\parallel {\bf c}$
(Fig.~2c) polarizations (note that Figs.~2a and 2b refer to the
14~$\mu$m thick sample, wile Fig.~2c represents the spectra of a
150~$\mu$m thick sample). The $a$-polarized FIR absorption continuum
extends to about 1500~cm$^{-1}$ and
consists of two bands centered at 300 and 1100\,cm
$^{-1}$~\cite{nav2,nav3,dam}.

Drastic changes occur in the spectra of NaV$_2$O$_5$ at the transition
temperature. First, the long-wavelength part of the broad-band
absorption in ${\bf
E}\parallel {\bf a}$ polarization decreases markedly, the sample becomes
more transparent (see Fig.~2a). The changes of transmittance near the
178~cm$^{-1}$ phonon line in ${\bf E}\parallel {\bf b}$ polarization
are caused by a sharp narrowing of this phonon below $T_c$. A special
study is necessary to clarify the question, whether the changes of
background transmittance observed in ${\bf E}\parallel {\bf c}$
polarization can be explained by a similar reason, namely, by the
changes of regular phonons 162, 179, and 282~cm$^{-1}$ below the
transition temperature, or there is a weak broad-band absorption that
diminishes below $T_c$. The spectra modeling (see below) has shown
that, indeed, a weak broad absorption band that bleaches below $T_c$
is present in ${\bf E}\parallel {\bf c}$ polarization. The measured
absorption coefficient at the frequency 115~cm$^{-1}$ is $\kappa
(\omega=115~{\rm cm}^{-1}) = 20~{\rm cm}^{-1}$. To compare, for the
${\bf E}\parallel {\bf a}$ polarized broad-band absorption $\kappa
(\omega=115~{\rm cm}^{-1}) = 400~{\rm cm}^{-1}$.

Second, the shape of
asymmetric peaks at 90 and 140~cm$^{-1}$ in ${\bf E}\parallel {\bf a}$
spectrum changes to symmetric. Third, numerous new spectral features
appear below $T_c$ in ${\bf E}\parallel {\bf a}$, ${\bf E}\parallel {\bf
b}$, and ${\bf E}\parallel {\bf c}$ spectra. The first two phenomena were
already discussed in our earlier publication~\cite{nav1}. We have
attributed a broad band in ${\bf E}\parallel {\bf a}$ polarization to
two-spinon absorption, the asymmetric line shape being due to Fano-type
resonance between lattice vibrations and a two-spinon continuum. The
opening of a gap in the magnetic excitation spectrum below $T_c$ leads to a
vanishing of a low-frequency continuum and to the appropriate
transformation of the line shapes. Subsequent structure
redetermination for the HT phase of
$\alpha'$-NaV$_2$O$_5$~\cite{smol,meet,vonschner,nav3}
does not, however, allow to ascribe  ${\bf E}\parallel {\bf a}$ polarized
spectral continuum to a pure two-spinon absorption. Symmetry
considerations show that for both HT- and LT-structure of
$\alpha'$-NaV$_2$O$_5$ the two-spinon (two-magnon) absorption is
allowed only in ${\bf E}\parallel {\bf c}$
polarization~\cite{moriya,valenti}.
Probably, a weak FIR absorption background observed in this
polarization results just from the two-spinon absorption. In that
case, the observed diminishing of this absorption in the
low-frequency region of ${\bf E} \parallel {\bf c}$ spectra would
correspond to the opening of a gap in the magnetic excitation
spectrum below $T_c$ and also one would expect to observe the
magnetic bound states, if any, just in ${\bf E}\parallel {\bf c}$
polarization. As for the ${\bf E}\parallel {\bf a}$ polarized
low-frequency continuum, it could be formed by two spinons and one
low-energy charge excitation, as has been proposed recently by
Khomskii et al. in their paper on the spin-isospin model of
$\alpha'$-NaV$_2$O$_5$~\cite{khom}.

In what follows, we concentrate our attention on the spectral lines that
appear below $T_c$.

\subsection{New spectral lines}

Numerous new spectral lines appear in transmittance spectra below
$T_c \simeq 34~$K and grow in intensity upon cooling the sample. For
the spectral region 50--350~cm$^{-1}$ they are marked by arrows in
Fig.~2. Figs.~3--5 show some examples of the ${\bf E}\parallel {\bf
a}$, ${\bf E}\parallel {\bf b}$, and ${\bf E}\parallel {\bf c}$
absorbance difference spectra $\alpha(T) - \alpha(37K$) for several
temperatures. Here, the absorbance is defined by $\alpha = -
\log_{10}{Tr/d}$, where $Tr$ is the transmittance of a $d$~cm thick
sample.

The temperature behavior of the majority of new lines is characterized by
an appreciable broadening and small (0.2--2~cm$^{-1}$) but clearly
detectable shifts near $T_c$ (see Figs.~\ref{fig6}, \ref{fig7}). It
should be mentioned that the vibrational lines that exist both below
and above $T_c$ also broaden and shift when approaching $T_c$ from
below (see insets of Fig.~\ref{fig6}). Positions, widths, and
oscillator strengths of the observed spectral lines that appear below
$T_c$ are listed in Table~\ref{tab1}.

These parameters were obtained from the following fitting procedure.
The experimental transmittance spectrum at normal incidence was
approximated by the expression~\cite{born}
\begin{equation}
Tr = \left|\frac{4 \sqrt{\varepsilon}}
{(\sqrt{\varepsilon}+1)^2 e^{-i \beta d} -
(\sqrt{\varepsilon}-1)^2 e^{i \beta d} }\right|^2,
\label{Tr}
\end{equation}
\begin{displaymath}
\beta= 2 \pi \omega \sqrt{\varepsilon}
\nonumber
\end{displaymath}
in the case of clearly visible interference pattern, or by the
expression
\begin{equation}
Tr = (1- {\cal R})^2 e^{- \kappa d}
\label{TrR}
\end{equation}
in the absence of interference within the sample. Here
\begin{equation}
{\cal R}=\left|\frac{\sqrt{\varepsilon}-1}
{\sqrt{\varepsilon}+1}\right|^2
\label{R}
\end{equation}
is the reflectivity of one surface,
\begin{equation}
\kappa = 4 \pi \omega Im \sqrt{\varepsilon}
\label{alpha}
\end{equation}
is the absorption coefficient. The dielectric function $\varepsilon
(\omega)$ was represented as a sum of contributions from $N$
noninteracting oscillators
\begin{equation}
\varepsilon = \varepsilon_{\infty}+\sum_{j=1}^N \frac{4\pi
f_j \omega_j^2} {\omega_j^2-\omega^2-i\gamma_j\omega},
\label{eps}
\end{equation}
where $\varepsilon_{\infty}$ is the high-frequency dielectric
constant, $\omega_j, \gamma_j$, and $f_j$ are the frequency, the damping
constant, and the oscillator strength of the $j$-th oscillator. First
of all, the calculated spectrum was fitted to the experimental one at
T=37\,K, just above the temperature of the phase transition, starting
from the room temperature values determined earlier in our
work~\cite{nav3}. The fitting to the low-temperature ($T\simeq$10\,K)
spectrum was performed in two steps. First, changes of the phonons
that exist both in HT- and LT-phases were taken into account and,
then, new oscillators were introduced.

It should be noted that we
could not study several spectral regions because of a too strong absorption
of either the sample or mylar windows of our cryostat. In particular, the
lines at 718~cm$^{-1}$ are only partially seen, being too close to a strong
absorption band of mylar. Results of the ${\bf E}\parallel {\bf c}$, ${\bf
k}\parallel {\bf a}$ measurements refer to the 150~$\mu$m thick sample.
We were able to detect the lowest frequency lines also in the 1.3~mm thick
sample. Their positions are indicated in brackets in Table~\ref{tab1}.

Fig.~8 shows the temperature dependence of the normalised oscillator
strength of the modes 101.4 and 111.7~cm$^{-1}$ (${\bf E}\parallel
{\bf b}$ polarization); 126.7~cm$^{-1}$ (${\bf E}\parallel
{\bf a}$); 70, 106.9, 124.7, 133.1, and 140~cm$^{-1}$ (${\bf
E}\parallel {\bf c}$).
The data on the 101.4 and 111.7~cm$^{-1}$ modes summarize a very
detailed study of two samples with the thicknesses $d_1=14~\mu$m and
$d_2=111~\mu$m.

The intensity of the 101.4, 126.7, and 140~cm$^{-1}$ modes follows the
same temperature dependence as the
intensity of new X-ray reflections corresponding to the doubling (a,
b) and quadrupling (c) of the lattice constants \cite{fujii} (also
shown in Fig.~8). This dependence is characterized by a steep rise
just below $T_c$ and a saturation below $\sim$ 22~K. The majority of
the observed new lines behave like that, while several ones (in
particular, the lines 91.2, 145.0, 145.7, 157.2, and 451.1~cm$^{-1}$
in ${\bf E}\parallel {\bf a}$, 111.7~cm$ ^{-1}$ in ${\bf E}\parallel
{\bf b}$, 70, 106.9, 126.7, and 133.1~cm$^{-1}$ in ${\bf E}\parallel
{\bf c}$) demonstrate a more slow growth of intensity below $T_c$ and
a saturation at lower temperatures. We could not measure reliably the
intensity dependence of the very weak feature at about 133~cm$^{-1}$
in ${\bf E}\parallel {\bf c}$ polarization but it seems to grow in
the slowest way among all the studied modes and to follow the same
temperature dependence as the spin-gap mode observed in optical ESR
measurements~\cite{luther} (also shown in Fig.~8).

\subsection{Factor-group analysis ($Fmm2$). Brillouin zone folding
for different proposed low-temperature structures.}

According to the recent X-ray diffraction
work~\cite{ludecke,deboer,thalm}, the
structure of the low-temperature phase of $\alpha'$-NaV$_2$O$_5$ is
face-centered, with $Fmm2$ space group on the $2a\times 2b\times 4c$
supercell.
The primitive cell corresponding to the translation vectors $(a,0,2c)$,
$(0,b,2c)$, and $(a,b,0)$ contains 64 atoms and, hence, there are 189
optical vibrational modes. Table~\ref{tab2} lists the Wyckoff
positions of atoms in the $Fmm2$ supercell \cite{ludecke} and in the
$Pmmn$ basic structure~\cite{smol}. The corresponding local
symmetries are also indicated. These positions yield the following
irreducible representations of the vibrational modes at the Brillouin
zone center of the $Fmm2$ LT-structure~\cite{porto}:
\begin{eqnarray*}
C_1      &: & 3A_1+3A_2+3B_1+3B_2 \\
C_{2v}   &: & A_1+B_1+B_2 \\
C_s^{xz} &: & 2A_1+A_2+2B_1+B_2 \\
C_s^{yz} &: & 2A_1+A_2+B_1+2B_2 \\
C_2      &: & A_1+A_2+2B_1+2B_2.
\end{eqnarray*}
Multiplying the representations given above by the number of different
positions of the appropriate symmetry, summing them, and subtracting
acoustic modes ($A_1+B_1+B_2$), we obtain the following optical vibrational
modes:
$$
\Gamma_{\scriptscriptstyle\rm NaV_2O_5}
      ^{\scriptscriptstyle\rm vib}(Fmm2)=
55A_1(aa,bb,cc;E\parallel c)+
36A_2(ab)+55B_1(ac;E\parallel a)+43B_2(bc;E\parallel b).
$$
153 modes ($A_1,B_1,B_2$) are both Raman and IR active, while $36A_2$ modes
are allowed only in Raman scattering.

More information on the transition-induced vibrational modes, in
particular, on the distribution of their frequencies, can be obtained
from the analysis of the BZ folding at the structural phase
transition. We have constructed the Brillouin zones for the $Pmmn$
HT-structure (it is shown in Fig.~9), for the $Fmm2$ LT-structure,
and also for the other proposed LT-structures and analysed what points
of the BZ of the paraphase  fold into the new zone center and thus
become optically active. Results of this analysis are represented in
Table~\ref{tab3}.

In the case of $Fmm2$ low-temperature structure, the two $Z$-points
$(0,0,\pm\frac12)$ of the paraphase and the eight $Q$-points of the type
$(\frac12,\frac12,\frac14)$ deliver 48 singlets $16A_1+6A_2+16B_1+10B_2$
(compatible with ($8Z_1+8Z_2)+(3Z_3+3Z_4)+(8Z_5+8Z_6)+(5Z_7+5Z_8$)
irreducible
representations of the symmetry group of the $Z$-point) and, respectively,
24 quadruplets $12(2A_1+2A_2)+12(2B_1+2B_2)$ (compatible with $Q_1$ and $Q_2$
doublets of the $Q$-point) to the BZ center of the LT phase.  They join
$15A_1+6A_2+15B_1+9B_2$ singlet vibrational modes that originate from,
correspondingly,
$(7B_{1u}+8A_g)+(3A_u+3B_{1g})+(7B_{3u}+8B_{2g})+(4B_{2u}+5B_{3g})$
vibrations of the BZ center ($\Gamma $-point) of the paraphase.
Because of a degeneracy of the $Q$-point folded modes, the number of
additional frequencies observed below $T_c$ should be not so high as
it followed from a simple factor-group analysis (namely, 72 instead of
$189-45=144$).

In the case of the face-centered structures $Ccc2$, $C2/c$, or $C2$,
there are 128 atoms in the primitive cell and, thus, 381 optical
modes. In addition to the just mentioned BZ points $Z$ and $Q$, the
$\Lambda$, $S$, and $R$ points also fold into the zone center
delivering 96 additional doublets. A lattice distortion in the LT-phase
lifts a degeneracy of folded quadruplets or quadruplets and doublets
and causes their splitting. We shall consider the polarization
properties of the split components in the section~IVA.

\subsection{Lattice dynamics calculations}

We consider the lattice dynamics of $\alpha'$-NaV$_2$O$_5$ in the
framework of the rigid-ion model. The model incorporates long-range
Coulomb interactions and short-range interactions described by
Born-Mayer potentials. The model parameters were obtained in
Ref.~\cite{nav3} by fitting to the HT-phase zone-center frequencies
measured by optical techniques. In the present work, we use them to
calculate the phonon dispersion curves throughout the Brillouin zone
of the paraphase. As an example, Fig.~10 displays the calculated
dispersion curves along the closed path $\Gamma (000) - \Lambda (00
\frac{1}{4}) - Z (00 \frac{1}{2}) -
R (\frac{1}{2}\frac{1}{2}\frac{1}{2}) -
Q (\frac{1}{2}\frac{1}{2}\frac{1}{4}) -
S (\frac{1}{2}\frac{1}{2}0) -
\Gamma (000)$ in the BZ of the high-temperature $Pmmn$ structure for
the two highest-frequency Davydov doublets $B_{3u} - B_{2g}$ and
$B_{1u} - A_g$ originating from the stretching vanadium-apical oxygen
vibration. We do not show here the rest of 48 phonon branches. Most of
them exhibit pronounced slopes and complicated interactions. Many
(but not all) of the dispersion curves along the $\Gamma$--$Z$
direction are rather flat, reflecting more weak forces between layers
in comparison with those inside a layer. However, only 12 out of 45
optical branches show a dispersion less than 6~cm$^{-1}$ (6~cm$^{-1}$
is a maximum line-width among the observed folded modes), and so the
phonon spectrum of the LT-phase cannot be described in the framework
of one layer, as has been proposed in Ref.~\cite{konst}.

Our lattice dynamical model neglects electronic polarization and does
not take into account three-body forces that strongly affect the
frequencies of bending vibrations. Because of that, there are several
large discrepancies (50--80~cm$^{-1}$) between the calculated and
observed zone-center phonon frequencies of the high-temperature
$Pmmn$ structure~\cite{nav3} and one cannot expect a good quantitative
description of phonon frequencies throughout the Brillouin zone. (It
should be mentioned, however, that the sound velocities determined
from the slope of the acoustic branches along the ladders (the $b$-axis) at
$k=0$,
$v=(\frac{d\omega}{dk})_{k=0}$,
$v_l=7.0\cdot10^5$\,cm/s,
$v_t^{(1)}=4.6\cdot10^5$\,cm/s,
$v_t^{(2)}=3.3\cdot10^5$\,cm/s, are in good agreement with the
experimentally measured values
$v_l=6.5\cdot10^5$\,cm/s,
$v_t^{(1)}=4.2\cdot10^5$\,cm/s,
$v_t^{(2)}=2.6\cdot10^5$\,cm/s~\cite{luthi}.)

Though being crude, this simple rigid-ion model gives possibility to
analyze, at least qualitatively, the main properties of the folded phonon
modes in the low-temperature phase of $\alpha'$-NaV$_2$O$_5$.  In
particular, to follow the temperature shifts of the normal mode frequencies
at temperatures near $T_c$, we have performed calculations in the framework
of $Fmm2$ LT-structure, using the atomic coordinates from the
structural work~\cite{ludecke} and taking into account the charge
redistribution. Three cases have been considered, namely, (i) lattice
deformation ignored, redistributed charges of vanadium ions in the
modulated ladders, V$^{4.5\pm\delta}$ with $\delta =0.05$; (ii) lattice
deformation ten times smaller than measured by X-ray diffraction at 15~K in
Ref.~\cite {ludecke}, $\delta =0.05$; (iii) the same as (ii) but
$\delta=0.1$. These calculations involved the numerical diagonalization of
the dynamical matrix with dimensions $192\times192$ that was
constructed using the same non-Coulomb force constants as in the paraphase.
The main results of calculations are as follows.

1) The soft acoustic mode at the $(0,0,\frac 12)$ point of the BZ of
the paraphase~\cite{nav3} becomes unstable under
redistribution of the vanadium charge density alone, not taking into
account the lattice distortion.

2) Both lattice distortion and charge
disproportion cause normal mode frequency shifts and splitting of
quadruplets (into $2A_1+2A_2$ or $2B_1+2B_2$ doublets),
due to induced changes of the dynamical
matrix. For different modes, displacements of atoms towards their new
equilibrium positions and charge redistribution have effects upon
frequencies either in the same or in the opposite direction. The
shifts of ``hard'' singlet modes and splitting of quadruplets amount
to 0.01--3\% of their frequency.

\section{Discussion}

\subsection{Splitting of the folded phonon modes}

First of all, we put the question whether it is possible to explain
the number and polarization properties of the additional
low-temperature modes of $\alpha'$-NaV$_2$O$_5$ supposing all of them
to be vibrational folded modes of a dimerized lattice.
Fourfold degeneracy of modes folded from the BZ $Q$-point at the phase
transition determines some important peculiarities of the
low-temperature vibrational spectrum of $\alpha'$-NaV$_2$O$_5$. We
shall discuss them assuming, first, the $Fmm2$
group for the low-temperature structure~\cite{ludecke,deboer,thalm}.
In this case, $Q_2$-quadruplets should manifest themselves as
$2B_1+2B_2$ pairs of doublets having close frequencies in ${\bf E}
\parallel {\bf a}$ and ${\bf E} \parallel {\bf b}$ FIR spectra. For
the two narrow additional low-temperature modes, namely, for the
lines at about 101.5 and 127~cm$^{-1}$ we really observe a gradual
splitting into such doublets upon lowering the temperature (see
Figs.~\ref{fig3},\ref{fig4},\ref{fig7}). We suppose that the observed
additional FIR lines with coincident or nearly coincident frequencies
in ${\bf E} \parallel {\bf a}$ and ${\bf E} \parallel {\bf b}$
spectra correspond to unresolved $2B_1+2B_2$ doublets originating
from $Q_2$ quadruplets. We place them into the lower right part of
Table~\ref{tab4}, together with Raman data from the literature. Note,
that the frequency differences between 199.0~cm$^{-1}$ (${\bf E}
\parallel {\bf a}$) and 199.5~cm$^{-1}$ (${\bf E} \parallel {\bf b}$)
lines and between 959.7~cm$^{-1}$ (${\bf E} \parallel {\bf a}$) and
959.2~cm$^{-1}$ (${\bf E} \parallel {\bf b}$) lines are beyond the
error limit of our high resolution measurements.

$Q_1$-quadruplets should split into $2A_1+2A_2$ pairs of doublets.
$A_1$ modes are observed in ${\bf E} \parallel {\bf c}$ polarization,
while $A_2$ modes are silent in the IR spectra (but allowed in the
(ab) Raman spectra). All the ${\bf E} \parallel {\bf c}$ modes are
rather broad and we never observed a reliably resolved doublet
structure. However,
the shape of the lowest frequency mode near 70~cm$^{-1}$ (${\bf E}
\parallel {\bf c}$) suggests a doublet structure (see Fig.~11). We
have performed a fitting of the measured spectrum by the expressions
(2)--(5) assuming a contribution from (i) one oscillator and (ii) two
oscillators. Fig.~11 clearly demonstrates a much better result in the
case of two oscillators. A doublet structure of the corresponding
Raman $A_1$-line has been reported in Ref.~\cite{konst2}. $A_1$ and
$A_2$ Raman modes with coincident or close frequencies are listed in
the lower left part of Table~\ref{tab4}. So, the lower part of
Table~\ref{tab4} contains
the LT modes that could originate from the BZ $Q$-point of the
paraphase. In that case, 7 out of 12 $Q_1$ and 10 out of 12 $Q_2$
quadruplets are observed. Upper part of Table~\ref{tab4} lists the
rest of the observed additional LT modes from our measurements and
from the literature. They could come from the Z-point, so that 10 out
of 16 $A_1$, 3 out of 6 $A_2$, 11 out of 16 $B_1$, and 3 out of 10 $B_2$
modes folded from the Z-point are observed. Raman and FIR modes that
coincide in frequency within the absolute precision of Raman
measurements (typically, $\pm 2$~cm$^{-1}$) are put into the same line of
Table~\ref{tab4}.

Thus, the number and polarization properties of the additional modes
that appear below the temperature of the phase transition in
$\alpha'$-NaV$_2$O$_5$ do not contradict their assignment to folded modes
of the distorted $Fmm2$ structure.

Now we compare in more detail the
experimental results to the results of lattice dynamics calculations for
the frequency range displayed in Fig.~10.  Table~\ref{tab5} lists the
calculated frequencies of the $\Gamma$-point TO and LO modes and Z- and
$Q$-point modes relevant to the zone folding in the case of $Fmm2$
LT-structure. All the experimentally observed modes in this frequency range
can be reasonably ascribed to the modes of the $Fmm2$ structure (see
columns 3 and 4 of Table~\ref{tab5}). The modes 948 ($A_1$) and
948~cm$^{-1}$ ($A_2$) with coincident frequencies in $A_1$ and $A_2$
Raman spectra and two clearly different infrared modes 959.7 ($B_1$)
and 959.2~cm$^{-1}$ ($B_2$) in ${\bf E} \parallel {\bf a}$ and
${\bf E} \parallel {\bf b}$ polarizations should
correspond to unresolved $2A_1$ and $2A_2$ and, respectively, $2B_1$
and $2B_2$ doublets folded from the $Q$-point. According to our lattice
dynamics calculations (for the case (iii) of the Section~IIIC),
$Q_1$ (969~cm$^{-1}$) and $Q_2$ (973~cm$^{-1}$) quadruplets (see
Fig.~10) split into $2A_1$ (967.8--970.3~cm$^{-1}$) and $2A_2$
(969.3--969.4~cm$^{-1}$) and, respectively, $2B_1$
(974.8--977.0~cm$^{-1}$) and $2B_2$ (973.5--973.6~cm$^{-1}$)
doublets. These splittings are within the limits of the linewidths.

If the space group for the LT-phase of $\alpha'$-NaV$_2$O$_5$ were a
subgroup of $Fmm2$ (e.g., $Ccc2$~\cite{ohama2,palstra}), 48 additional
doublets $16 \times 2 A_1 + 6 \times 2 A_2 + 16 \times 2 B_1 + 10
\times 2 B_2$ from the $\Lambda$ point would appear independently in
the spectra of different symmetries and also one would expect
additional 96 singlets $24 (A_1 + A_2)$ and $24 (B_1 + B_2)$ with
close frequencies in $A_1$ and $A_2$ and $B_1$ and $B_2$ spectra,
originating from the folded $S$- and $R$-doublets. Domains due to oblique
charge ordering pattern possible for a zigzag charge order in every
ladder in the space group $Ccc2$~\cite{ohama2,dam} would not affect the
direction of the orthorhombic a- and b-axes and, thus, polarization
properties of vibrational modes. Experimental data on the new
LT-modes do not contradict both $Fmm2$ and $Ccc2$ symmetries.
However, while in the first case one has to admit that 27 new
frequencies remained unobserved, in the latter one this quantity
grows up to 123. In both cases, close or coincident frequencies in
$A_1$ and $A_2$ and $B_1$ and $B_2$ spectra are naturally explained
by their origin from the folded BZ $Q$-point ($Fmm2$) or Q, $\Lambda$,
S, and $R$-points ($Ccc2$). A possible monoclinic distortion (space groups
$C2$~\cite{palstra} or $C2/c$~\cite{konst}) would violate the ${\bf
E} \parallel {\bf a}$ or ${\bf E} \parallel {\bf b}$ FIR selection
rules, as well as Raman selection rules. Probably, the growing contribution
of the $A_{1g}$ modes 530 and 304~cm$^{-1}$ to the (ab) Raman spectrum for
the temperatures below $\sim100$\,K observed in Ref.~\cite{fischer2} is
just because of such a distortion.

At the first glance, LT modes fit the $C2/c$ centrosymmetric group
proposed in Ref.~\cite{konst}. In particular, in the case of $C2/c$
group $Q_2$-quadruplets would split into $2 B_u ({\bf E} \parallel
{\bf a},  {\bf E} \parallel{\bf b}) + 2 B_g (ac, bc)$ doublets. Such a
scenario seems to describe better the experimental data (see the
lower right part of Table~\ref{tab4}). However, the frequency
differences between the observed FIR and Raman modes are within the
precision of Raman measurements.

To summarize, it is not possible to chose between the proposed space
symmetry groups for the low-temperature crystal structure of
$\alpha'$-NaV$_2$O$_5$ on purely spectroscopic grounds. The number
and polarization properties of all the observed new LT modes in
$\alpha'$-NaV$_2$O$_5$ can be reasonably described within the
conception of folded vibrational modes of the dimerized
$Fmm2$ crystal structure. Observed FIR doublets and close
frequencies in different polarizations are naturally explained by
their origin from the $Q$-point modes folded into the zone center.
Prior to analyse the temperature dependences of frequencies and
widths of new modes and to discuss alternative interpretations, we
note that published FIR \cite{nav1,dsmirnov1,dsmirnov2,nav3,dam} and
Raman \cite{weiden,kuroe,lemmens,fischer2,konst,konst2,konst3} data
agree with each other, within the precision of measurements (typically,
$\pm0.5$~cm$^{-1}$ for FIR and $\pm1.5$~cm$^{-1}$ for Raman
measurements).

\subsection{Magnetic bound states?}

As we mentioned in Section~IIIA, three of the observed new
low-temperature modes, namely, modes near 70, 107, and 133~cm$^{-1}$
in ${\bf E} \parallel {\bf c}$ polarization, have different
frequencies in different samples (see Table~\ref{tab4}). These modes
and the mode near 125~cm$^{-1}$ (${\bf E} \parallel {\bf c}$)
demonstrate unusually large broadening, shift, and loss of intensity
upon heating (see Figs.~6,~8). All these modes were observed also in
Raman scattering (see Table~\ref{tab4}), exhibiting the same
properties as their infrared counterparts.

In Refs.~\cite{kuroe,konst2}, Raman
modes near 65 and 132~cm$^{-1}$ have been assigned to one- and
two-magnon scattering processes, respectively.  The frequency
64--70~cm$^{-1}$ of the lowest-frequency sharp feature in FIR and Raman
spectra below $T_c$ corresponds well to the size of the gap
$\Delta_0=8.13$~meV$=65.6$~cm$^{-1}$ at ${\bf k=0}$ obtained from
submillimeter ESR measurements \cite{luther}. The transitions between
the singlet ground state and the excited triplet state are forbidden
($\Delta S=1$). The Dzyaloshinsky-Moria (DM) interaction with DM
vector along the $c$-axis breaks the selection rule for ESR (and FIR)
${\bf E} \parallel {\bf c}$ transitions in
$\alpha'$-NaV$_2$O$_5$~\cite{luther,valenti}. Recently, it has
been shown theoretically \cite{valenti} that for
$\alpha'$-NaV$_2$O$_5$ the DM interaction results in finite Raman
cross-section for both one- and two-magnon scattering processes. A
single-magnon Raman line of this type should show no splitting in an
external magnetic field parallel to the $c$-axis  but should split
into two branches for a field perpendicular to the $c$-axis. The
authors of Ref. \cite{fischer2} communicated that the 67~cm$^{-1}$
Raman line neither shifts nor splits nor even broadens in a magnetic
field up to 7~T. They have attributed this line and the 107 and
134~cm$^{-1}$ Raman modes to magnetic singlet bound states, on the
basis of (i) the temperature dependence of the intensity, frequency
shift, and linewidth, and (ii) the frequency dependence on a sample
quality. All the three modes rapidly loose their intensity,
soften and broaden on approaching $T_c$ from below
\cite{lemmens,fischer2}. It is worth mentioning that very recently
two more Raman modes with similar properties have been reported
(namely, 86 and 126~cm$^{-1}$ modes in (ab) geometry)~\cite{konst3}.
To our opinion, the interpretation of the 67, 107, and 134~cm$^{-1}$
modes as magnetic bound states does not follow unambiguously from the
above mentioned arguments.

One of the main arguments in favor of considering  67, 107, and
134~cm$^{-1}$ Raman modes as magnetic bound states was a slow
increase of  their intensity upon cooling. For comparison, the
intensities of the folded phonon Raman modes at 202, 246, and
948~cm$^{-1}$ rose sharply below $T_c$ and then saturated.
However, our measurements show that many of the
transition-induced modes demonstrate a slow growth of intensity upon
cooling below $T_c$. Among them, the considered mode near
107~cm$^{-1}$ (${\bf E}\parallel {\bf c}$) and the folded phonon
mode 111.7~cm$^{-1}$ (${\bf E}\parallel {\bf b}$) behave in the same
way (see Fig.~8).

Different temperature dependences of the type $(T_c - T)^{2n \beta}$
(here $\beta$ is the critical index of the order parameter; $n=1, 2,
3$) for the intensities, frequencies and widths of the folded modes
in the vicinity of a structural phase transition have been reported
and analysed by Petzelt and Dvo\v r\'ak~\cite{dvorak}, taking into
account anharmonic couplings between modes. In
$\alpha'$-NaV$_2$O$_5$, as we show below, a mere charge disproportion
may lead to the $(T_c - T)^{2n \beta}, n > 1$, intensity dependence
for a folded phonon mode below $T_c$, even in the harmonic approximation.

Intensity of a FIR
spectral line is proportional to a square of the electric dipole moment of
a primitive cell induced by the respective vibrational mode. The dipole
moment may be represented as a sum of products of effective atomic charges
and corresponding displacements of Bravais  sublattices. A slow
intensity dependence may be expected when the main contribution to the
dipole moment comes from the charge disproportion terms that involve
displacements renormalized by the  order parameter. Let us consider,
for example, the $z$-component of the electric dipole moment due to
displacements of vanadium ions that transform according to the $A_1$
representation of the factor  group of the LT phase. The primitive
cell of the $Fmm2$ structure contains
eight V$^{4.5+}$ ions that form two rectangular plaquetts (fragments of
nonmodulated ladders) in the neighboring $ab$-layers,
four V$^{4+}$, and four V$^{5+}$ ions on the rungs of the neighboring (left
and right) modulated ladders (see Fig.~\ref{fig1}). Normal modes
that correspond to $Z_1$, $Z_2$, and $Q_1^1$ representations
involve antiphase translations of V$^{4.5+}$ plaquetts along the $c$-axis
and antiphase displacements of V$^{5+}$ and V$^{4+}$ ions along the same
axis. At $T_c$, all the displacements have equal amplitudes, vanadium
ions have
equal charges, and the corresponding electric dipole moment is zero. At
$T<T_c$, nonzero dipole moment appears due to different amplitudes of
the plaquette translations and different amplitudes of V$^{4.5-\delta}$ and
V$^{4.5+\delta}$ ions. Our calculations confirmed that, for some low
frequency modes involving large amplitudes of vanadium ions vibrations, the
effective electric dipole moment is determined mainly by the terms
containing the charge disproportion $\delta$ and the differences
between ion amplitudes at the temperatures $T$ and $T_c$.  These terms
provide the $(T_c - T)^{4 \beta}$ dependence of the FIR absorption.
The same consideration is applicable to the Raman spectra where
intensities are determined by electric polarizabilities of lattice
modes.

Thus, a slow growth of the mode intensity below $T_c$ is not a
unique property of the spin-gap related excitations only. Futhermore,
if we compare the intensities of the FIR new modes measured in this
work with the integrated intensity of the spin-gap mode observed in
optical ESR measurements~\cite{luther}, it is evident that only the
133~cm$^{-1}$ mode follows the same temperature dependence as the
gap-mode (see Fig.~8).

Now in Fig.~12 we compare the normalized magnetic gap energy, as
observed in neutron scattering experiments~\cite{fujii}, and the
normalized frequencies of the 70 and 106.9~cm$^{-1}$ FIR modes. The
latter change much more slowly on approaching T$_c$ than the magnetic
gap. We could not measure the frequency of the weak
133~cm$^{-1}$ FIR mode below 24~K but the corresponding Raman mode
practically does not shift within the interval of $T/T_c$ from 0.2 to
0.88, as one can see from the spectra of Ref.~\cite{konst3}. Thus,
the temperature dependence of the peak positions for the lines
near 70, 107, and 124.7~cm$^{-1}$ differ significantly from the
dependence for the singlet-triplet gap, $\Delta~(T)$,
while the frequency of the Raman
peak near 30~cm$^{-1}$ corresponding to a magnetic bound state in
CuGeO$_3$ followed the same temperature dependence as
$\Delta~(T)$~\cite{kuroe2}.

To summarize, the comparison between the temperature-dependent
intensities of the transition-induced FIR modes and of the spin-gap
mode show that the mode 133~cm$^{-1}$ is the only one that could be
considered as a magnetic bound state. However, the temperature
dependence of its frequency contradicts this hypotesis.
Thus, the assignment of the 70, 107, and 133~cm$^{-1}$ modes to magnetic
bound states seems to be wrong.

As an alternative interpretation we would
propose folded phonon modes that strongly interact with charge and spin
excitations. The interaction renormalizes the frequencies of phonons,
causes mode broadening and softening near $T_c$. A sodium defficiency
strongly influences magnetic (and charge) states which leads to a
strong change of interaction parameters. This circumstance could
explain the dependence of 70, 106, and 133~cm$^{-1}$ mode frequencies
on a sample quality. Our earlier observation on the $x$-dependent
phonon frequencies in Na$_{1-x}$V$_2$O$_5$~\cite{nav3} supports such
an assumption.
We observed a giant frequency shift of 30~cm$^{-1}$
for one of the (aa) Raman modes. Namely, the 447~cm$^{-1}$ mode
($x=0$) moved to 477~cm$^{-1}$ for $x=0.15$.
Simultaneously, the
maximum of a broad-band scattering moved in the opposite direction,
from 632~cm$^{-1}$ for $x=0$ to 562~cm$^{-1}$ for $x=0.15$. It was
naturally to explain this change of the frequency difference between
two Raman bands by a change of interaction between the corresponding
excitations~\cite{nav3}. Though the exact nature of the broad band in
the (aa) Raman geometry is not yet clear, it must be connected with
low-frequency electronic excitations on the ladder
rungs~\cite{fischer2,khom}.

The interaction between spin, charge, and
lattice degrees of freedom is an intrinsic property of
$\alpha'$-NaV$_2$O$_5$.
In sodium vanadate one $d$-electron belongs to two vanadium atoms
which gives an additional electronic degree of freedom. The electron
hopping amplitude along a ladder rung is significantly higher than the
amplitudes of hopping along a ladder or between different
ladders~\cite{smol,horsch}. Zigzag charge ordering at low
temperatures can be described as an antiferroelectric ordering of
electric dipoles on the rungs of vanadium ladders~\cite{khom}. As
each electron carries the spin (S=1/2),
charge and spin degrees of freedom are closely connected. To describe
this situation, Mostovoy and Khomskii proposed a spin-isospin model
and used it to show that the antiferroelectric charge ordering in the
system of vanadium ladders opens a spin gap~\cite{most,khom}. They
argued that in $\alpha'$-NaV$_2$O$_5$ the quasi-one-dimensional spin
system is strongly coupled to a low-energy antiferroelectric mode
(isospin mode) of the excitonic type and the observed charge ordering
corresponds to the softening of this mode. The softening of the
isospin excitation occurs at the wave vector of the superlattice
structure appearing in the ordered phase~\cite{khom}, that is, at the
$Q$-point ($\frac{1}{2}$$\frac{1}{2}$$\frac{1}{4}$). Low-energy
vibrational excitations of the corresponding symmetry should strongly
interact with the isospin mode. To understand a possible mechanism of
such an interaction we refer to Fig.~13 showing atomic displacements
for vanadium atoms in the lowest-frequency $A_1$ mode (at
136~cm$^{-1}$, according to our calculations) folded from the BZ
$Q$-point into the zone center. Vibrations are mainly along the
$c$-axis, vanadium atoms at the two ends of a rung move in antiphase,
that is when the left atom approaches the apical oxygen, the right
one moves away from it. Such a motion stimulates vanadium charge
redistribution (V$^{4.5 \pm \delta}$)~\cite{loosdr} that corresponds
just to a static picture of charge ordering.

By inspecting the experimental results summarized in Table~\ref{tab1}
we notice that the lowest frequency $A_1$-symmetry modes (${\bf
E}\parallel {\bf c}$) are about an order of magnitude broader than
the lowest-frequency $B_1$ (${\bf E}\parallel {\bf a}$) and $B_2$
(${\bf E}\parallel {\bf b}$) modes even in the limit $T \rightarrow
0$. They have the ratio $\gamma/ \omega \simeq 2 \div 6 \cdot 10^{-2}$
in comparison with $1 \div 3 \cdot 10^{-3}$ for $B_1$ and $B_2$ modes.
Such a broadening could result from a decay of the low-frequency
vibrations into isospin and spin excitations, due to the interaction
with them. High-resolution neutron inelastic scattering
experiments on the spin excitations of $\alpha'$-NaV$_2$O$_5$
revealed the two branches associated with distinct energy
gaps~\cite{yosihama,grenier}. At the zone boundary
$(\frac{1}{2} \frac{1}{2}0)$ (as well as at the zone center) these
gaps constitute 68.6~cm$^{-1}$~\cite{yosihama} (67~cm$^{-1}$
according to Ref.~\cite{grenier}) and 91.1~cm$^{-1}$~\cite{yosihama}
(87.8~cm$^{-1}$~\cite{grenier}).
Keeping in mind the energy and momentum conservation laws, the
following decay channels are possible: direct interaction with the zone
center magnon or a decay into two zone boundary isospin excitations
(for the 70~cm$^{-1}$ mode), a decay into a zone boundary
magnon (66 or 90~cm$^{-1}$) and the isospin
excitation (for the 106 or 125~cm$^{-1}$ modes). Such a process
corresponds to a hopping of an electron along the ladder rung and a
simultaneous spin-flip. In that case, the zone boundary isospin gap
would constitute of about 35~cm$^{-1}$, in reasonable agreement with
the theoretical estimates~\cite{khom}.

\subsection{In search for a soft mode}

The lattice dynamics calculations for the high-temperature phase of
$\alpha'$-NaV$_2$O$_5$ predicted a soft mode behavior of the
transverse acoustic mode at the BZ boundary with ${\bf k}=
(00\frac{1}{2})$. It was found to become unstable under charge
redistribution below $T_c$. So, one could expect a doubling of the
unit cell along the $c$-axis, as a precursor of the subsequent
magnetic ordering with a dimerization along the a-, b-, and
c-directions. We have to mention that two subsequent phase
transitions within the temperature interval of 1~K have been
communicated in Ref.~\cite{koppen}.

The discussed soft mode corresponds to the irreducible representation
Z$_5$ of the BZ $(00\frac{1}{2})$ point and it could be observed as a
folded FIR mode in ${\bf E} \parallel {\bf a}$ polarization. We have
thoroughly studied low-temperature
$a$-polarized FIR transmittance spectra of
$1\mbox{mm}\times3\mbox{mm}\times111\mu\mbox{m}$
sample of $\alpha'$-NaV$_2$O$_5$ in the spectral range
20--138~cm$^{-1}$ but failed to find a soft mode. Possibly, it has a
frequency lower than 20~cm$^{-1}$ at 6~K or its oscillator strength
is too small. Another possibility would be that this soft mode is
simply a property of our rather crude model of lattice dynamics.

The ${\bf E} \parallel {\bf c}$ polarized 70 and 107~cm$^{-1}$ modes
demonstrate the strongest observed softening toward $T_c$.
But this softening is only of about 3~$\%$ and is, probably, connected
with the interaction of lattice modes with
magnetic and charge excitations, as discussed in the previous
section. It looks like the true soft mode of the phase transition in
$\alpha'$-NaV$_2$O$_5$ is of electronic nature (low-energy
antiferroelectric (isospin) mode~\cite{khom}).

\section{Conclusions}

In summary, we report the temperature-dependent polarized
high-resolution FIR transmission spectra of $\alpha'$-NaV$_2$O$_5$.
The comparison with the results of neutron scattering and optical ESR
measurements and the theoretical analysis of the properties of
vibrational modes in a charge ordered state lead us to the conclusion
that all the transition-induced new modes of $\alpha'$-NaV$_2$O$_5$
can be completely described in terms of folded modes of a dimerized
lattice.

Based on the present results, the widely accepted interpretation of
several lowest-frequency new modes as magnetic bound states seems
highly unlikely. As an alternative
interpretation of 70, 106, 125, and 133~cm$^{-1}$
spectral lines we suggest low-frequency lattice vibrations of an
appropriate symmetry that strongly interact with charge and spin
excitations at the BZ boundary of the high-temperature phase and fold into
the zone center at the phase transition into a dimerized state.

{\bf Note:} When this paper already was with editors be became aware
of the work~\cite{jap} where ${\bf E}\parallel {\bf c}$ transmittance
of $\alpha'$-NaV$_2$O$_5$ was measured and low-frequency
transition-induced modes were reported to appear at 68, 106, and
124~cm$^{-1}$  ($T=4.2$~K) and not to change under the magnetic field
up to 15~T.

\section*{Acknowledgement}

Useful discussions with A.~N.~Vasil'ev, A.~I.~Smirnov, and G.~N.~Zhizhin
are acknowledged. We thank D.~M.~Shkrabo for modeling several
spectra. We are grateful to P. van Loosdrecht for critically reading
the manuscript. This work was supported by INTAS, grant
No.~99--0155 and by the Russian Foundation for Basic Research, grant
No.~01--02--16329.

\newpage
\begin{table}[h]
\caption{Parameters at $T=9$ K of the FIR absorption lines
activated by the phase transition in $\alpha'$-NaV$_2$O$_5$. The
frequencies of ${\bf E}\parallel {\bf c}$ modes in the sample 2 are
given in brackets. Here $\omega_j$ is the mode frequency,
$\gamma_j$ is the damping constant, $f_j$ is the oscillator strength.}
\begin{center}
\begin{tabular}{ccc|ccc||lcc}
& ${\bf E}\parallel {\bf a}$ &  &
& ${\bf E}\parallel {\bf b}$ &  &
& ${\bf E}\parallel {\bf c}$ &  \\
$\omega_j$, cm$^{-1}$ & $\gamma_j$, cm$^{-1}$ & $10^4 \cdot f_j$ &
$\omega_j$, cm$^{-1}$ & $\gamma_j$, cm$^{-1}$ & $10^4 \cdot f_j$ &
$\omega_j$, cm$^{-1}$ & $\gamma_j$, cm$^{-1}$ & $10^4 \cdot f_j$ \\
\hline
91.2  & 0.17 & 4.7 & 101.4 & 0.20 & 5.2 & 70.0 (67.5) & 4.5 & 33 \\
101.4 & 0.24 & 8.0 & 111.7 & 0.24 & 1.2 & 106.9 (106.2) & 2.3 & 10 \\
101.7 & 0.16 & 3.6 & 126.8 & 0.31 & 3.4 & 124.7 (124.7) & 1.0 & 1.6 \\
125.6 & 0.24 & 0.4 & 127.5 & 0.33 & 9.0 & 133.1 (131.6) & 2.5 & 0.5\\
126.8 & 0.33 & 13.0 & 199.5 & 1.3 & 2.5 & 140.0 & 0.7 & 1.3  \\
127.6 & 0.30 & 4.8 & 234.2 & 2.0 & 2.6  & 148.2 & 0.6 & 0.7\\
145.0 & 0.46 & 3.2 & 324.6 & 3.9 & 5.2  & 256.5 & 2.6 & 0.05 \\
145.7 & 0.43 & 7.6 & 410.3 & 2.0 & 1.5  & 327.4 & 6.0 & 0.17  \\
147.9 & 0.5 & 26   & 959.2 & 1.3 & 0.06 &  &  &   \\
157.2 & 0.3 & 0.8 &  &  &  &  &  &  \\
199.0 & 1.8 & 46.2  &  &  &  &  &  &  \\
410.4 & 4.3 & 26  &  &  &  &  &  &  \\
451.1 & 2.9 & 3.7 &  &  &  &  &  &  \\
959.7 & 1.5 & 0.4 &  &  &  &  &  &  \\
\end{tabular}
\end{center}
\label{tab1}
\end{table}

\begin{table}[h]
\caption{Wyckoff positions of atoms in the elementary cells of
$\alpha'$-NaV$_2$O$_5$ above and below $T_c$. Figures before chemical
symbols denote numbers of atoms of the same kind.}
\begin{center}
\begin{tabular}{cccc|cccc}
\multicolumn{4}{c|}{$T>T_c$, $Pmmn$~\cite{smol}} &
\multicolumn{4}{c}{$T<T_c$, $Fmm2$~\cite{ludecke}} \\
\multicolumn{2}{c}{Atom} & Wyckoff position & Site symmetry &
\multicolumn{2}{c}{Atom} & Wyckoff position & Site symmetry \\
\hline\hline
1 & V & $4f$ & $C_s^{xz}$ & 2 & V1 & $16e$ & $C_1$ \\
&  &  &  & 4 & V2 & $8d$ & $C_s^{xz}$ \\ \hline
1 & Na & $2b$ & $C_{2v}$ & 4 & Na1 & $4a$ & $C_{2v}$ \\
&  &  &  & 2 & Na2 & $8b$ & $C_2$ \\ \hline
1 & O1 & $2a$ & $C_{2v}$ & 2 & O11 & $8c$ & $C_s^{yz}$ \\
&  &  &  & 2 & O12 & $8d$ & $C_s^{xz}$ \\ \hline
1 & O2 & $4f$ & $C_s^{xz}$ & 2 & O21 & $16e$ & $C_1$ \\
&  &  &  & 4 & O22 & $8d$ & $C_s^{xz}$ \\ \hline
1 & O3 & $4f$ & $C_s^{xz}$ & 2 & O31 & $16e$ & $C_1$ \\
&  &  &  & 4 & O32 & $8d$ & $C_s^{xz}$
\end{tabular}
\end{center}
\label{tab2}
\end{table}

\begin{table}[hbtp]
\caption{Number of atoms in a primitive cell n, number of BZ center
optical vibrational modes N; points in the BZ of the paraphase
that fold into the zone center and corresponding optical modes in the
cases of different proposed space groups for the distorted structure.
m is the number of additional frequencies just below $T_c$.}
\begin{tabular}{lllll|l}
Space group & n & N & BZ points & Optical modes & m \\
\hline
& & & & & \\
$Fmm2$~\cite{ludecke,deboer,thalm} & 64 &  189 & $\Gamma$, Z, Q
& $45 \Gamma + 48Z + 24 \times 4Q$ & 72 \\
& & & & & \\
\hline
& & & & & \\
$Ccc2$~\cite{ohama2,palstra}, & 128 & 381 & $\Gamma$, Z, Q,
$\Lambda$, S, R & $45 \Gamma + 48Z + 24 \times4Q$  & \\
$C2/c$~\cite{konst}, & & & & $+ 48 \times2 \Lambda + 24 \times2S$ & 168 \\
and $C2$~\cite{palstra} & & & & $+ 24 \times2R$ & \\
& & & & & \\
\end{tabular}
\label{tab3}
\end{table}

\newpage
\begin{table}[hbtp]
\caption{Additional low-temperature IR and Raman modes of
$\alpha'$-NaV$_2$O$_5$. Figures in brackets refer to the second
sample used in this work or in Ref.~\protect\cite{fischer2}.}
\begin{tabular}{ll|l|ll|ll}
\multicolumn{2}{c|}{$A_1$} & $A_2$
& \multicolumn{2}{c|}{$B_1$} & \multicolumn{2}{c}{$B_2$} \\
\hline
IR & Raman & Raman & IR & Raman & IR & Raman \\
${\bf E} \parallel c$ & (aa), (bb), (cc) & (ab) &
${\bf E} \parallel a$ & (ac) & ${\bf E} \parallel b$ & (bc) \\
\hline
107 (106) & 107 (103)$^a$ & 86$^c$ & 91.2 &  & 111.7 &  \\
& & & & 116$^e$ & & \\
140.0 & 140$^b$  & & 125.6 &  & 234.2  &  \\
& 164$^a$ & 159$^b$, 158$^c$ & 145.1 &  &  & 354$^e$ \\
$*$ & 202$^a$ & 422$^{a,d}$ & 145.7 &  &  &  \\
256 &  &  & 147.9 &  &  &  \\
& 296$^b$ &  & 157.2 & 157$^f$ &  & \\
314 &  &  & $*$ & 306$^f$ &  &  \\
$*$ & 394$^a$ &  & 374$^g$ &  &  &  \\
$*$ & 650$^a$ &   & 451.1 &  &  &  \\
$*$ & 692$^a$ &   & 787.3 &  &  &  \\
&  &   & [969$^h$]  &  &  &    \\
\hline
70.0 (67.5) & 67 (64)$^a$ & 67 (64)$^a$ & 101.4 &  & 101.4 &  \\
 &  &  & 101.7 & 102$^f$ &  &  \\
124.7 (124.7) & 127$^b$, 123$^i$ & 127$^b$, 126$^c$ & 126.8 & & 126.8 &  \\
133.1 (131.6) & 134 (131)$^a$ & 134 (131)$^a$ & 127.6 & 128$^f$ & 127.5 &  \\
 &  &  & 199.0 & 200$^f$ & 199.5 &  \\
148.2 & 151$^a$ & 149$^b$, 148$^c$ & $*$ & 323$^e$  & 324.6 & 323$^e$ \\
244 & 246$^a$ & 246$^{a,d}$ & 362$^g$ &  & 362$^g$ &  \\
327.4 & 325$^a$ & 332$^{a,d}$  & 410.3 & 408$^e$ & 410.3 & 408$^e$ \\
$*$ & 948$^a$ & 948$^a$ & $*$ & 433$^e$ & $*$ & 433$^e$ \\
&  &  & 717$^g$ & 720$^e$ & 718$^g$ & 720$^e$ \\
&  &  & 959.7 &  & 959.2 &  \\
\end{tabular}
\label{tab4}
\noindent
$^*$ Strong absorption due to HT-phonons in this frequency region \\
$^a$ Low-temperature Raman data from Ref.~\cite{fischer2} \\
$^b$ Weak feature in the low-temperature spectra of Ref.~\cite{fischer2} \\
$^c$ Low-temperature Raman data from Ref. \cite{konst3} \\
$^d$ Fano resonance \\
$^e$ Low-temperature Raman data from Ref.~\cite{konst} \\
$^f$ Weak feature in the low-temperature spectra of Ref.~\cite{konst}\\
$^g$ Low-temperature FIR data from Ref.~\cite{dam} \\
$^h$ Nonpolarized FIR transmittance data for a powdered sample from
Ref.~\cite{dsmirnov2} \\
$^i$ Weak feature in the low-temperature (aa) spectrum of
Ref.~\cite{konst3}. \\
\end{table}

\newpage

\begin{table}[hbtp]
\caption{Comparison between the calculated vibrational frequencies of
the $\Gamma$, Q, and Z-modes for the high-temperature $Pmmn$
structure and measured frequencies for the HT- and LT-phases of
$\alpha'$-NaV$_2$O$_5$. Corresponding irreducible representations for
the $Fmm2$ LT-structure are indicated.}
\begin{tabular}{l|lll|l}
$Pmmn$ & \multicolumn{2}{c}{calculated} & measured & $Fmm2$ \\
\hline
$\Gamma$ & $B_{3u}$ (TO-LO)& 955--957 & 939 (${\bf E} \parallel
a$)~\cite{nav3}  & $B_1$ \\
 & $B_{1u}$ (TO-LO)& 961--1036 & 955--1017 (${\bf E} \parallel
c$)~\cite{nav3}  & $A_1$ \\
 & $B_{2g}$ & 962 & 951 (ac)~\cite{nav3}  & $B_1$ \\
 & $A_g$ & 964 & 970 (cc, aa)~\cite{nav3}  & $A_1$ \\
\hline
Q & $Q_1$ & 969 & 948 (aa, bb)--948 (ab)~\cite{fischer2} & $2A_1+2A_2$  \\
  & $Q_2$ & 973 & 959.7 (${\bf E} \parallel a$)--959.2 (${\bf E}
\parallel b$) & $2B_1+2B_2$  \\
\hline
Z & Z$_5$ & 985 & $969^a$~\cite{dsmirnov2} &  $B_1$ \\
  & Z$_6$ & 1003 & $*$ &  $B_1$ \\
  & Z$_1$ & 1018 & $*$ &  $A_1$ \\
  & Z$_5$ & 1024 & $*$ &  $A_1$ \\
\end{tabular}
\label{tab5}
\noindent
$^*$ Strong absorption due to HT-phonons in this frequency region \\
$^a$ Observed in unpolarized transmittance spectra of a powder sample \\
\end{table}

\newpage
\section*{Figure captions}

\begin{figure}[tbp]
\caption{The structure of the low-temperature phase of
$\alpha'$-NaV$_2$O$_5$. (a) Grey pyramids incorporate V$^{4.5+}$
ions. Black (white) pyramids are bigger (smaller) in size and are occupied
by the V$^{4+}$ (V$^{5+}$) ions. Balls represent Na atoms. (b) The
$ab$-projection of the one V--O layer. Black (white) circles stand for
V$^{4+}$ (V$^{5+}$), while the grey ones represent V$^{4.5+}$. Arrows
indicate the direction of displacements at the phase transition into
the LT-phase.}
\label{fig1}
\end{figure}

\begin{figure}[tbp]
\caption{Transmittance spectra of $\alpha'$-NaV$_2$O$_5$ at $T
\approx 40$~K$ > T_c$ (dashed line) and $T=8$~K$ < T_c$ (solid line)
in ${\bf E}\parallel{\bf a}$, ${\bf E}\parallel{\bf b}$, and ${\bf E}
\parallel{\bf c}$ polarizations of the incident light with ${\bf
k}\parallel{\bf c}$ (a,b) and ${\bf k}\parallel{\bf a}$ (c). The
sample thicknesses were 14~$\mu$m (a,b) and 150~$\mu$m (c). Arrows
show new low-temperature modes below $T_c$.}
\label{fig2}
\end{figure}

\begin{figure}[tbp]
\caption{Absorbance difference spectra in ${\bf E}\parallel{\bf a}$
polarization for different temperatures  showing the growth of new
spectral lines below $T_c$. Spectra are vertically shifted for the
sake of clarity.}
\label{fig3}
\end{figure}

\begin{figure}[tbp]
\caption{Absorbance difference spectra in ${\bf E}\parallel{\bf b}$
polarization: transition-induced modes with frequencies below
230~cm$^{-1}$.}
\label{fig4}
\end{figure}

\begin{figure}[tbp]
\caption{Absorbance difference spectra in ${\bf E}\parallel{\bf c}$
polarization. An asterisk marks weak features due to a leakage of
${\bf E}\parallel{\bf a}$ polarization. A weak intrinsic $c$-polarized mode
near 133~cm$^{-1}$ is shown by an arrow.} \label{fig5}
\end{figure}

\begin{figure}[tbp]
\caption{Temperature dependences of the normalized shifts and
linewidths of transition-induced modes. The insets show temperature
dependences for the two modes from the high-temperature phase of
$\alpha'$-NaV$_2$O$_5$.}
\label{fig6}
\end{figure}

\begin{figure}[tbp]
\caption{Peak position of selected transition-induced modes at
different temperatures.}
\label{fig7}
\end{figure}

\begin{figure}[tbp]
\caption{Temperature dependence of the normalized oscillator strengths
$f_T/f_{8{\rm K}}$ for several transition-induced modes of
$\alpha'$-NaV$_2$O$_5$. Lines are guides for the eye.
Intensities of the X-ray
reflections from Ref.~\protect\cite{fujii} are shown by grey symbols.
Asterisks represent the normalized integrated intensity of the
spin-gap mode observed in optical ESR, as estimated from the data of
Ref.~\protect\cite{luther}.}
\label{fig8}
\end{figure}

\begin{figure}
\caption{Brillouin zone of the $Pmmn$ HT-structure of
$\alpha'$-NaV$_2$O$_5$.}
\label{fig9}
\end{figure}

\begin{figure}[tbp]
\caption{Calculated phonon dispersion for the high-temperature $Pmmn$
structure of $\alpha'$-NaV$_2$O$_5$.}
\label{fig10}
\end{figure}

\begin{figure}[pbt]
\caption{Lowest-frequency transition-induced mode near 70~cm$^{-1}$
in  ${\bf E}\parallel{\bf c}$ transmittance spectrum of the
150~$\mu$m thick sample, corrected for the additional absorption of
the epoxy layer (open circles). The dashed line was calculated using
Eqs.~(2)--(5) and one-oscillator approximation
($\omega=70.0$~cm$^{-1}$, $\gamma=4.5$~cm$^{-1}$, $f=3.3 \cdot
10^{-3}$). The solid line represents the fit result for the two
oscillators ($\omega_1 = 68.75$~cm$^{-1}$, $\omega_2=70.9$~cm$^{-1}$;
$\gamma_1=2.2$~cm$^{-1}$, $\gamma_2=3.2$~cm$^{-1}$; $f_1=1.2 \cdot
10^{-3}$, $f_2=1.85 \cdot 10^{-3}$).}
\label{fig11}
\end{figure}

\begin{figure}[pbt]
\caption{Comparison between temperature dependences of normalized
energies (frequencies) $\hbar \omega_T/\hbar \omega_{8~K}$ of the
transition-induced modes near 70 and 107~cm$^{-1}$ and normalized
magnetic gap $\Delta_T/\Delta_{7~K}$ form inelastic neutron
scattering measurements~\protect\cite{fujii} (black circles).}
\label{fig12}
\end{figure}

\begin{figure}[pbt]
\caption{Picture of atomic displacements of vanadium atoms for the
lowest-frequency $A_1$ mode ($Fmm2$) folded from the $Q$-point of
($Pmmn$) high-temperature structure. (a) one ab-layer. Only vanadium
atoms are shown. Displacements along the positive (negative)
direction of the c-axis are indicated by points (crosses). (b)
Vanadium atoms 1 and 2 at the opposite ends of a ladder rung,
together with oxygen at the middle of a rung and two apical oxygens.
When V2 approaches apical oxygen, V1 goes moves away from it, which
stimulates an asymmetric charge distribution along the rung.}
\label{fig13}
\end{figure}


\begin{references}

\bibitem{c1}  M. Isobe and Y. Ueda, J. Phys. Soc. Jap. {\bf 65}, 1178
(1996).

\bibitem{smol}  H. Smolinski, C. Gros, W. Weber, U. Peuchert, G. Roth, M.
Weiden, and C. Geibel, Phys. Rev. Lett. {\bf 80}, 5164 (1998).

\bibitem{meet}  A. Meetsma, J. L. de Boer, A. Damascelli, T. T. M. Palstra,
J. Jegoudez, and A. Revcolevschi, Acta Cryst. C {\bf 54} 1558 (1998).

\bibitem{vonschner}  H. G. von Schnering, Y. Grin, M. Kaupp, M. Somer, R.
K. Kremer, O. Jepsen, T. Chatterji, and M. Weiden, Z. Kristallogr. {\bf
213}, 246 (1998).

\bibitem{nav3}  M. N. Popova, A. B. Sushkov, S. A. Golubchik, B. N. Mavrin,
V. N. Denisov, B. Z. Malkin, A. I.Iskhakova, M. Isobe, and Y. Ueda, ZhETF
{\bf 115}, 2170 (1999); [JETP {\bf 88}, 1186 (1999)].

\bibitem{ludecke}  J. L\"udecke, A. Jobst, S. van Smaalen, E. Morr\'e, C.
Geibel, and H. G. Krane, Phys. Rev. Lett. {\bf 82}, 3633 (1999).

\bibitem{ohama}  T. Ohama, H. Yasuoka, M. Isobe, and Y. Ueda,
Phys. Rev. B {\bf 59}, 3299 (1999).

\bibitem{horsch}  P. Horsch and F. Mack, Eur. Phys. J. B {\bf 5}, 367
(1998).

\bibitem{weiden}  M. Weiden, R. Hauptmann, C. Geibel, F. Steglich, M.
Fisher, P. Lemmens, and G. Guntherodt, Z. Phys. B {\bf 103}, 1 (1997).

\bibitem{fujii}  Y. Fujii, H. Nakao, T. Yoshihama et al., J. Phys. Soc.
Jap.  {\bf 66}, 326 (1997).

\bibitem{nav5}  A. I. Smirnov, M. N. Popova, A. B. Sushkov, S. A.
Golubchik, D. I. Khomskii, M. V. Mostovoy, A. N. Vasil'ev, M. Isobe, and Y.
Ueda, Phys.  Rev. B 59, 14546 (1999).

\bibitem{luthi}  H. Schvenk, S. Zherlitsyn, B. L\"uthi, E. Morre, and C.
Geibel, Phys. Rev. B {\bf 60}, 9194 (1999).

\bibitem{x-anom} H. Nakao, K. Ohwada, N. Takesue, Y. Fujii, M. Isobe,
Y. Ueda, M. v. Zimmermann, J. P. Hill, D. Gibbs, J. C. Woicik,
I. Koyama, and Y. Murakami, Phys. Rev. Lett. {\bf 85}, 4349 (2000).

\bibitem{deboer}  J. L. de Boer, A. M. Meetsma, J. Baas, and T. T. M.
Palstra, Phys. Rev. Lett. {\bf 84}, 3962 (2000).

\bibitem{thalm} A. Bernert, T. Chatterji, P. Thalmeier, and P. Fulde,
cond-mat/0012327.

\bibitem{europhys}  S. van Smaalen and J. L\"udecke, Europhys. Lett. {\bf
49}, 250 (2000).

\bibitem{ohama2} T. Ohama, A. Goto, T. Shimizu, E. Ninomiya, H. Sawa,
M. Isobe, and Y. Ueda, J. Phys. Soc. Jap. {\bf 69}, 2751 (2000).

\bibitem{palstra} J. L. de Boer, G. Maris, A. M. Meetsma, J. Baas,
and T. T. M. Palstra, cond-mat/0008054.

\bibitem{konst}  M. J. Konstantinovi\'c, Z. V. Popovi\'c, A. N.
Vasil'ev, M. Isobe, and Y. Ueda, Solid State Commun. {\bf 112}, 397
(1999).

\bibitem{nav1}  M. N. Popova, A. B. Sushkov, A. N. Vasil'ev, M. Isobe, and
Y. Ueda, Pis'ma v ZhETF {\bf 65} 711 (1997) [JETP Lett. {\bf 65} 743
(1997)].

\bibitem{kuroe}  H. Kuroe, H. Seto, J. Sasaki, T. Sekine, M. Isobe, and Y.
Ueda, J. Phys. Soc. Japan {\bf 67}, 2881 (1998).

\bibitem{fischer} M. Fischer, P. Lemmens, G. G\"{u}ntherodt,
M. Weiden, R. Hauptmann, C. Geibel, and F. Steglich, Physica B {\bf
244}, 76 (1998).

\bibitem{dsmirnov1}  D. Smirnov, P. Millet, J. Leotin, D. Poilblanc, J.
Riera, D. Augier, and P. Hansen, Phys. Rev. B {\bf 57}, R11035 (1998).

\bibitem{dam2} A. Damascelli, D. van der Marel, M. Gr\"uninger, C.
Pressura, T. T. M. Palstra, J. Jegoudez, and A. Revcolevschi,
Phys. Rev. Lett. {\bf 81}, 918 (1998).

\bibitem{mism}  M. N. Popova, A. B. Sushkov, S. A. Golubchik, M. Isobe, and
Y. Ueda, Physica B {\bf 284--286}, 1617 (2000).

\bibitem{dsmirnov2}  D. Smirnov, J. Leotin, P. Millet, J. Jegoudez, and
A. Revcolevschi, Physica B {\bf 259--261}, 992 (1999).

\bibitem{lemmens}  P. Lemmens, M. Fischer, G. Els, G. G\"untherodt,
A. S. Mishchenko, M. Weiden, R. Hauptmann, C. Geibel, and F. Steglich,
Phys.  Rev. B {\bf 58}, 14159 (1998).

\bibitem{fischer2}  M. Fischer, P. Lemmens, G. Els, G. G\"untherodt, E. Ya.
Sherman, E. Morr\'e, C. Geibel, and F. Steglich, Phys. Rev. B {\bf 60},
7284 (1999).

\bibitem{konst3}  M. J. Konstantinovi\'c, J. C. Irwin, M. Isobe, and Y.
Ueda, cond-mat/0103453.

\bibitem{isobe}  M. Isobe, C. Kagami, and Y. Ueda, J. Crystal Growth {\bf
181}, 314 (1997).

\bibitem{nav2}  S. A. Golubchik, M. Isobe, A. N. Ivlev, B. N. Mavrin, M. N.
Popova, A. B. Sushkov, Y. Ueda, and A. N. Vasil'ev, J. Phys. Soc. Japan
{\bf 66}, 4042 (1997).

\bibitem{dam}  A. Damascelli, C. Presura, D. van der Marel, J. Jegoudez,
and A. Revcolevschi, Phys. Rev. B {\bf 61}, 2535 (2000).

\bibitem{moriya}  T. Moriya, J. Appl. Phys. {\bf 39}, 1042 (1968).

\bibitem{valenti}  R. Valenti, C. Gros, and W. Brenig, Phys. Rev. B
{\bf 62}, 14164 (2000); cond-mat/0006036.

\bibitem{khom} M. Mostovoy, J. Knoester, and D. Khomskii,
cond-mat/0009464.

\bibitem{born} M. Born, E. Wolf, Principles of optics, 2nd ed.
(Pergamon Press, Oxford, London, Edinburgh, New York, Paris,
Frankfurt, 1964).

\bibitem{luther}  S. Luther, H. Nojiri, M. Motokawa, M. Isobe, and Y.
Ueda, J. Phys. Soc. Jpn {\bf 67}, 3715 (1998); {\bf 69}, 2291 (2000).

\bibitem{porto}  D. L. Rousseau, R. P. Bauman, and S. P. S. Porto, J.
Raman Spectr. {\bf 10}, 253 (1981).

\bibitem{konst2}  M. J. Konstantinovi\'c, K. Ladavac, A. Beli\'c,
Z. V. Popovi\'c, A. N. Vasil'ev, M. Isobe, and Y. Ueda, J.
Phys.:Condens. Matter {\bf 11}, 2103 (1999).

\bibitem{dvorak}  J.~Petzelt and V.~Dvo\v r\'ak, J. Phys. C: Solid State
Phys. {\bf 9}, 1571 (1976).

\bibitem{kuroe2}  H. Kuroe, T. Sekine, M. Hase, Y. Sasago, K. Uchinokura,
H. Kojima, I. Tanaka, and Y. Shibuya, Phys. Rev. B {\bf 50}, 16468 (1994).

\bibitem{most}  M. V. Mostovoy and D. I. Khomskii, Solid State Commun. {\bf
113}, 159 (2000).

\bibitem{loosdr} E. Ya. Sherman, M. Fischer,  P. Lemmens, P. H. M. van
Loosdrecht, and G. G{\"u}ntherodt, Europhys. Lett. {\bf 48}, 648
(1999).

\bibitem{yosihama}  T. Yosihama, M. Nishi, K. Nakajima, K. Kakurai,
Y. Fujii, M. Isobe, C. Kagami, and Y. Ueda, J. Phys. Soc. Japan {\bf 67},
744 (1998).

\bibitem{grenier}  B. Grenier, O. Cepas, L. P. Regnault, J. E. Lorenso, T.
Ziman, J. P. Boucher, A. Hiess, T. Chatterji, J. Jegoudez, and A.
Revcolevschi, cond-mat/0007025.

\bibitem{koppen}  M. K\"oppen, D. Pankert, R. Hauptmann, M. Lang, M.
Weiden, C. Geibel, and F. Steglich, Phys. Rev. B {\bf 57}, 8466 (1998).

\bibitem{jap} K. Tahekana, T. Takamasu, G. Kido, M. Isobe, and Y.
Ueda, Physica B {\bf 294--295}, 79 (2001).


\end{references}
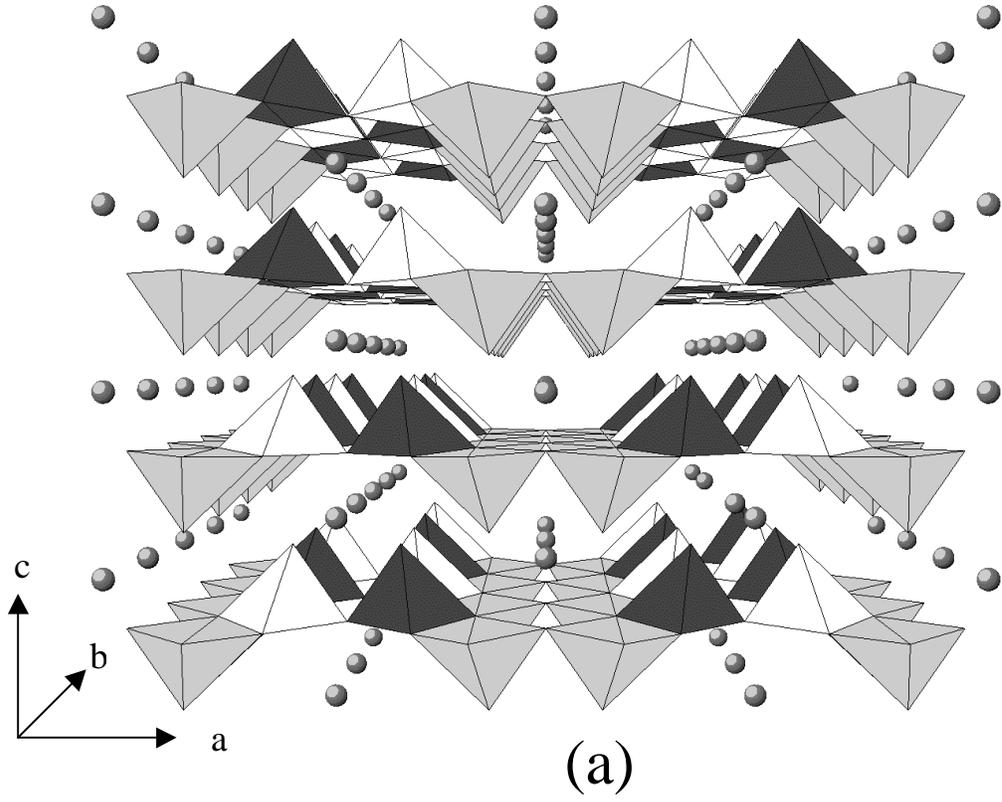
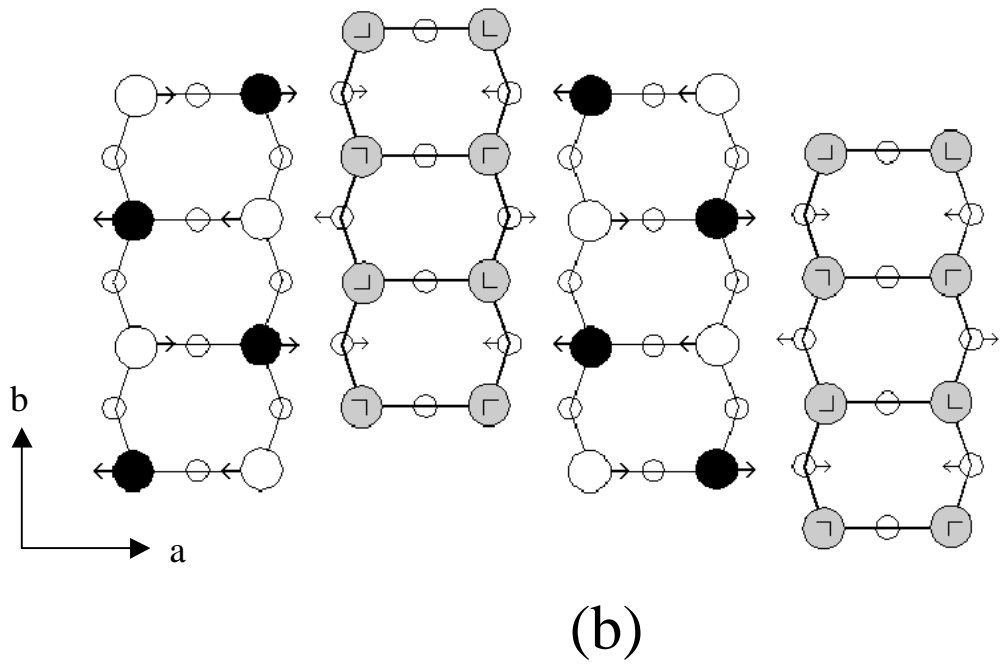
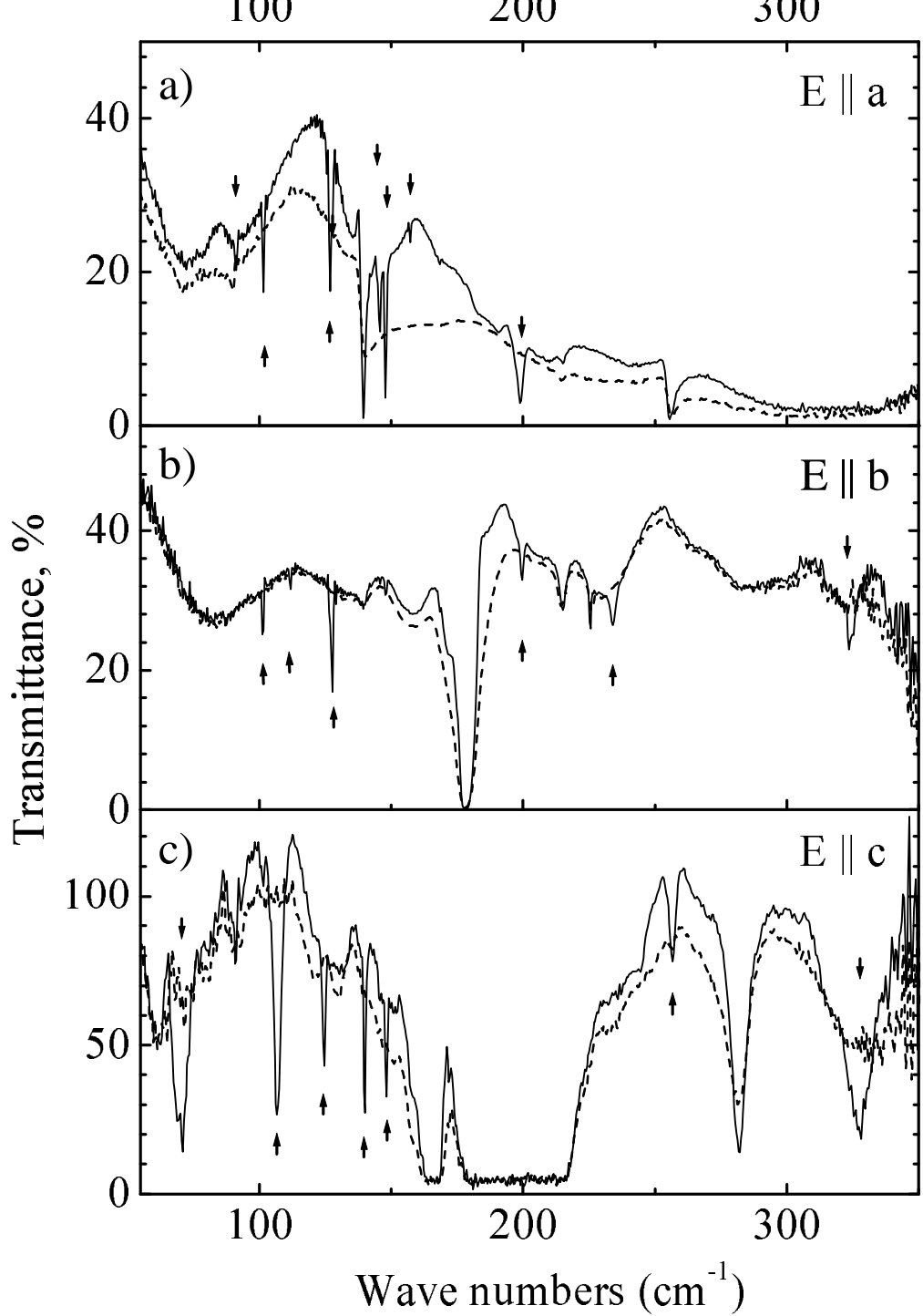
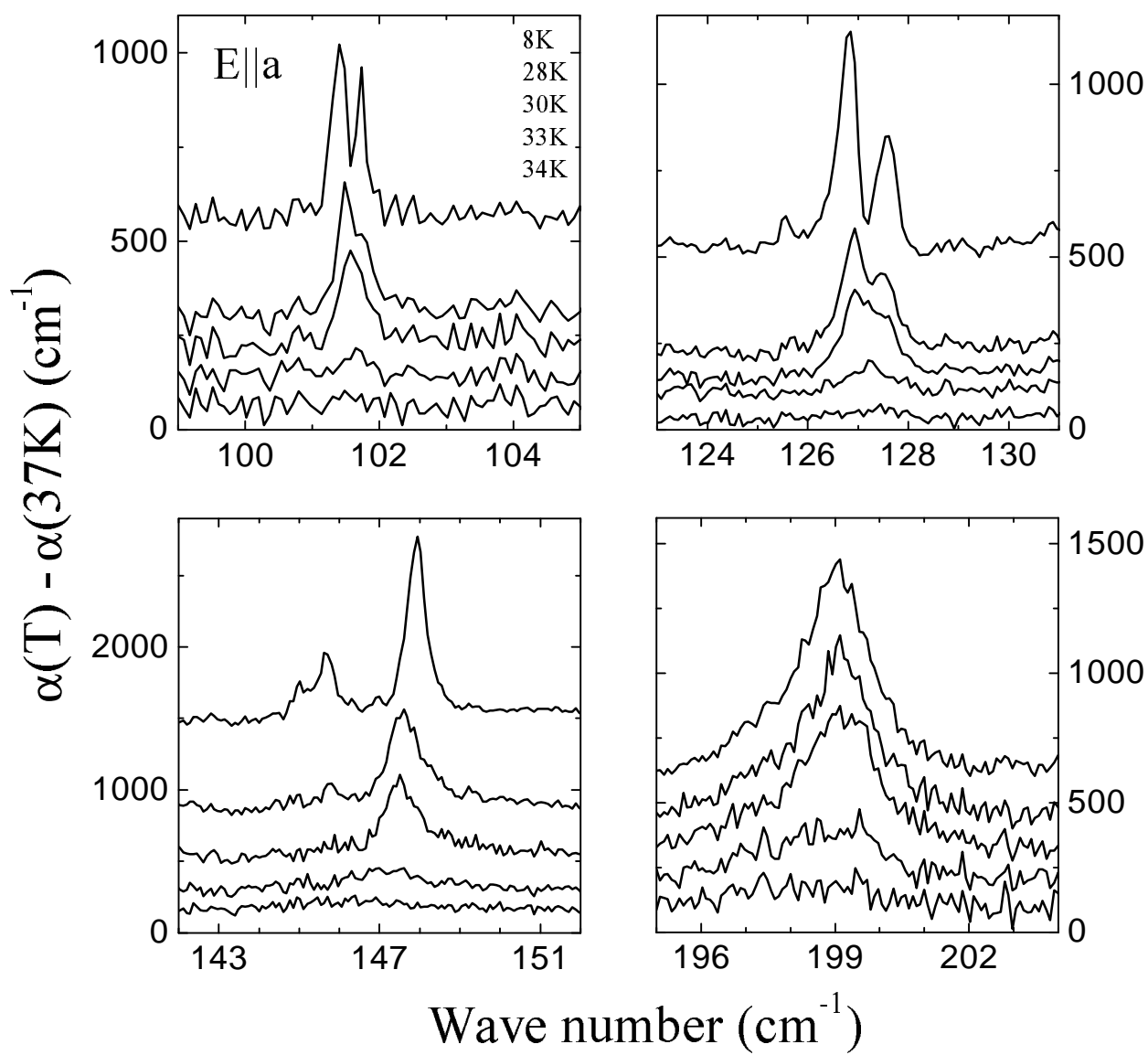
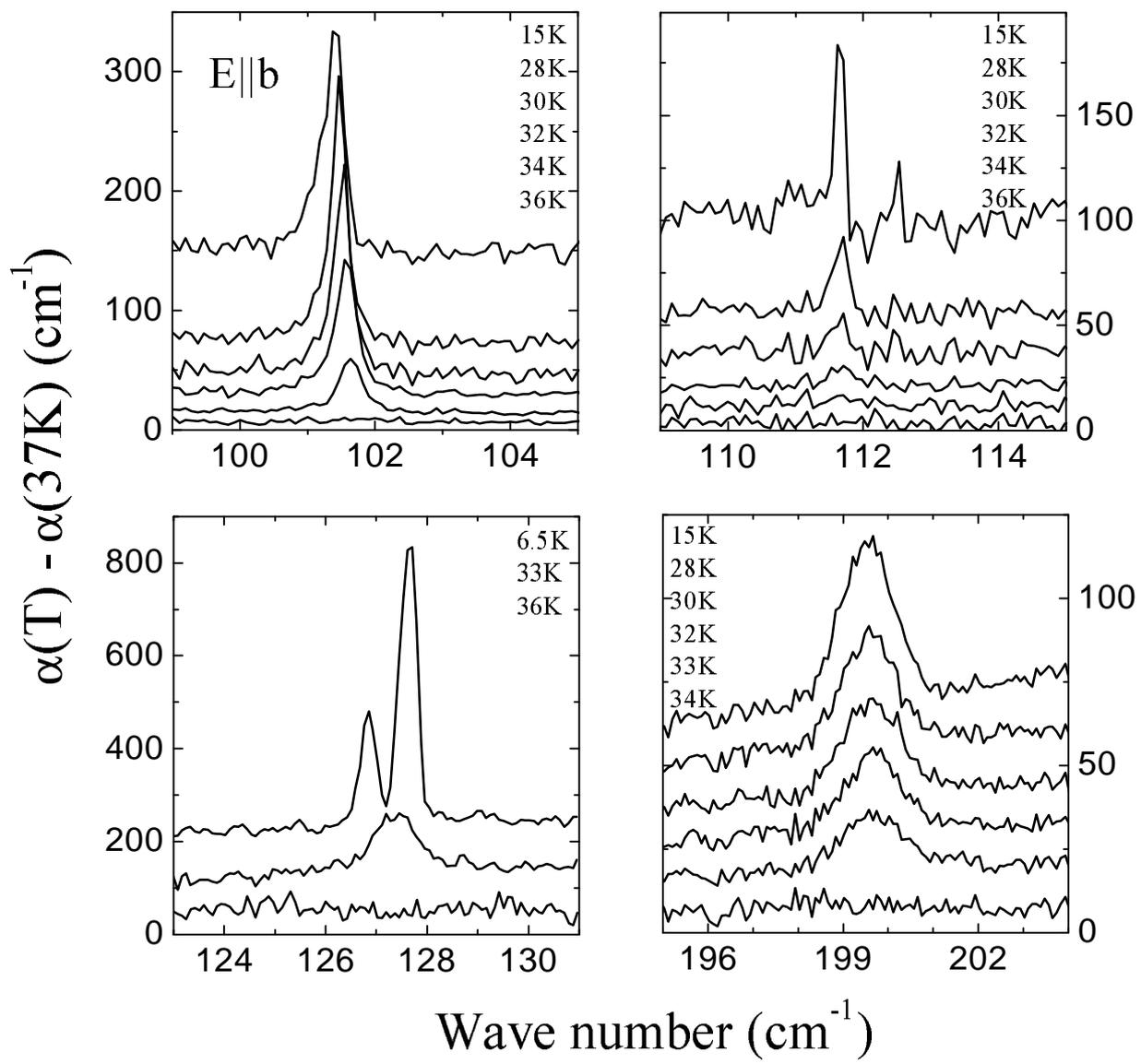
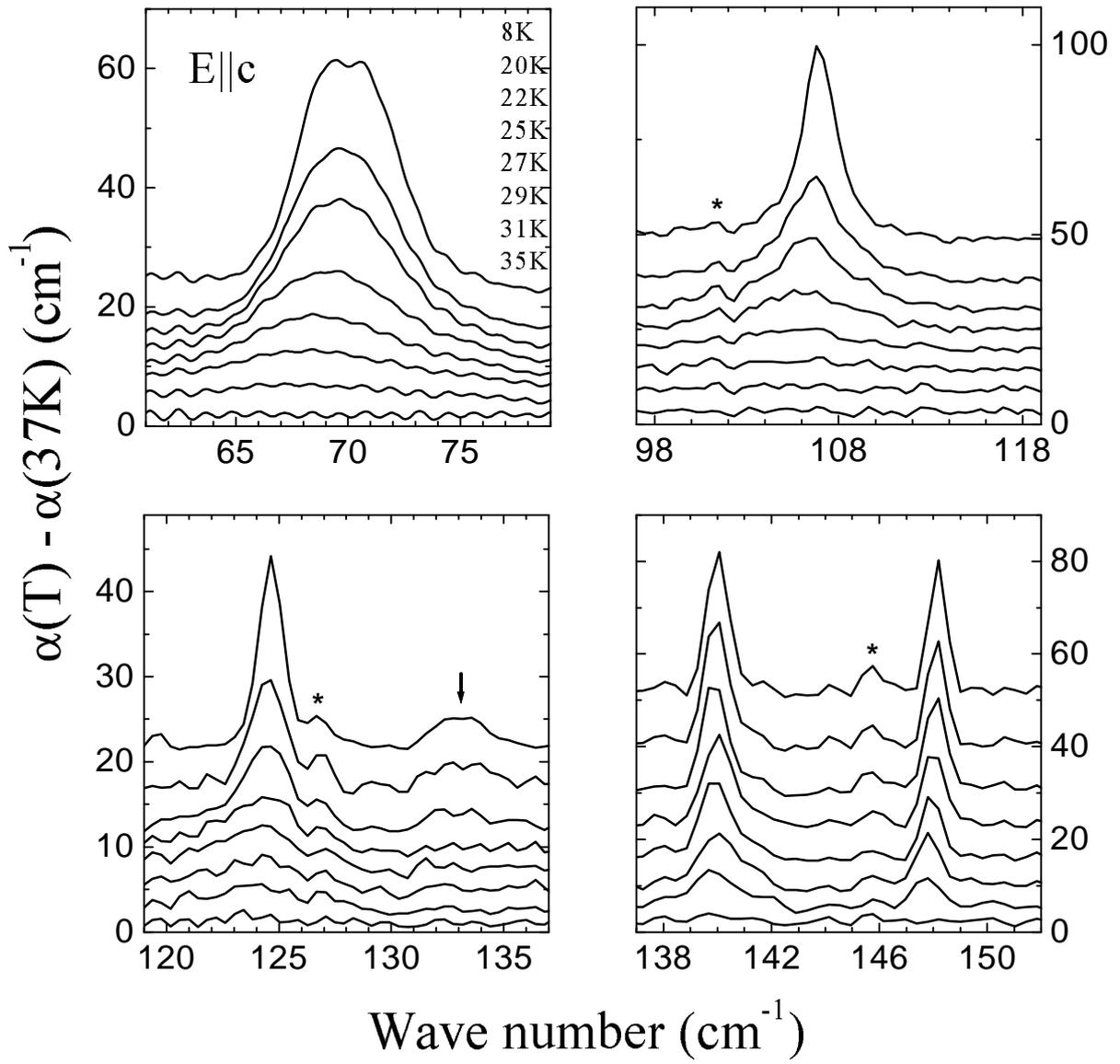
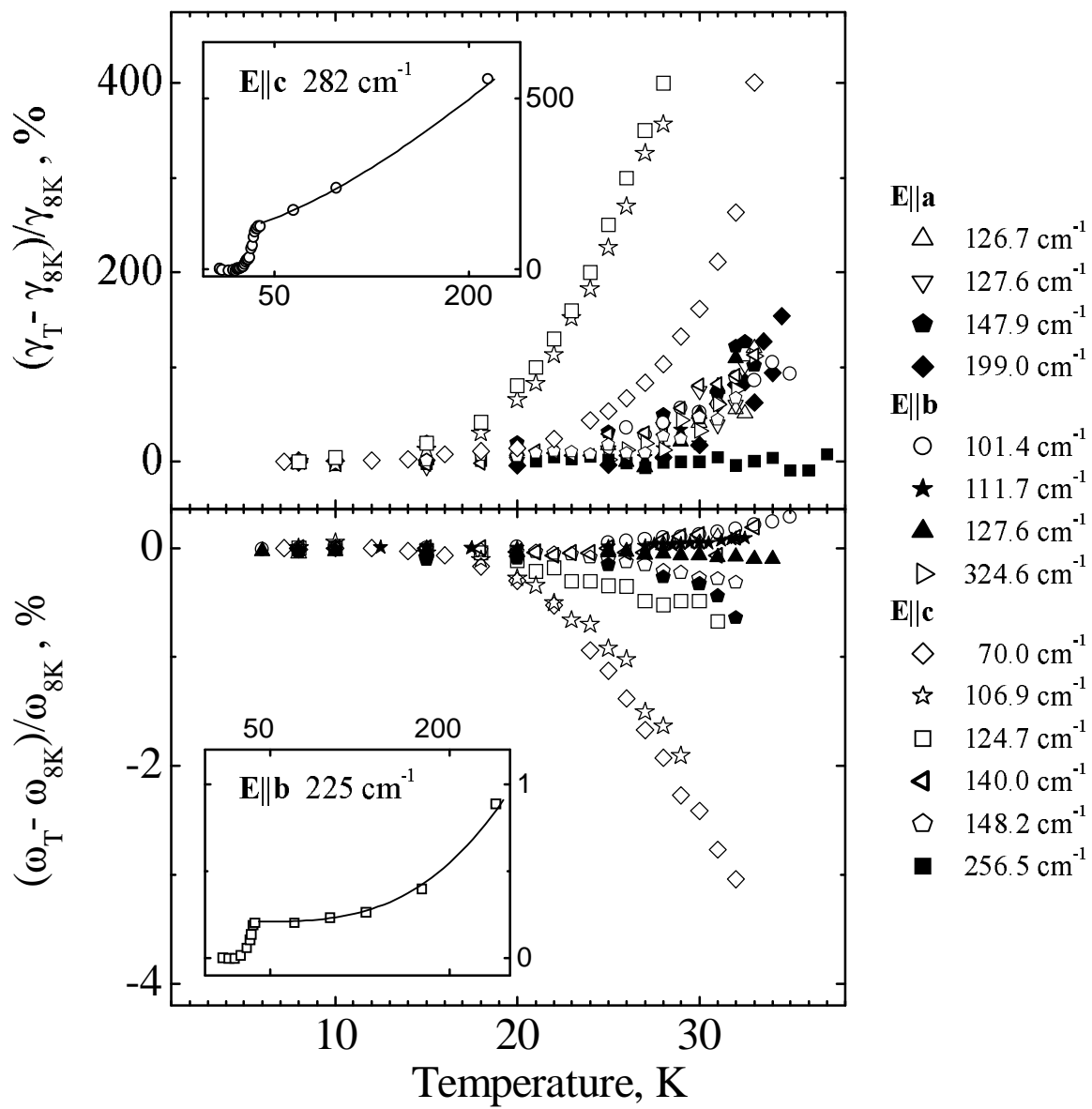
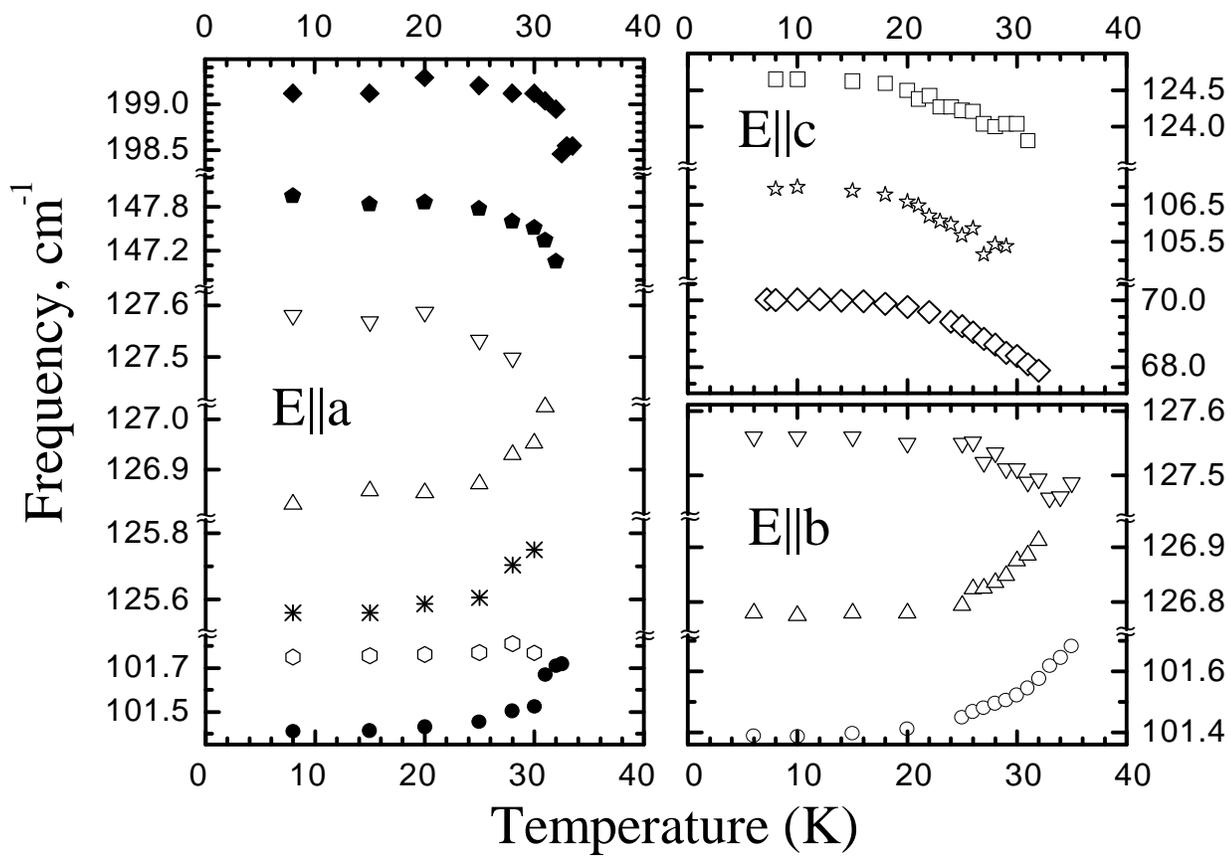
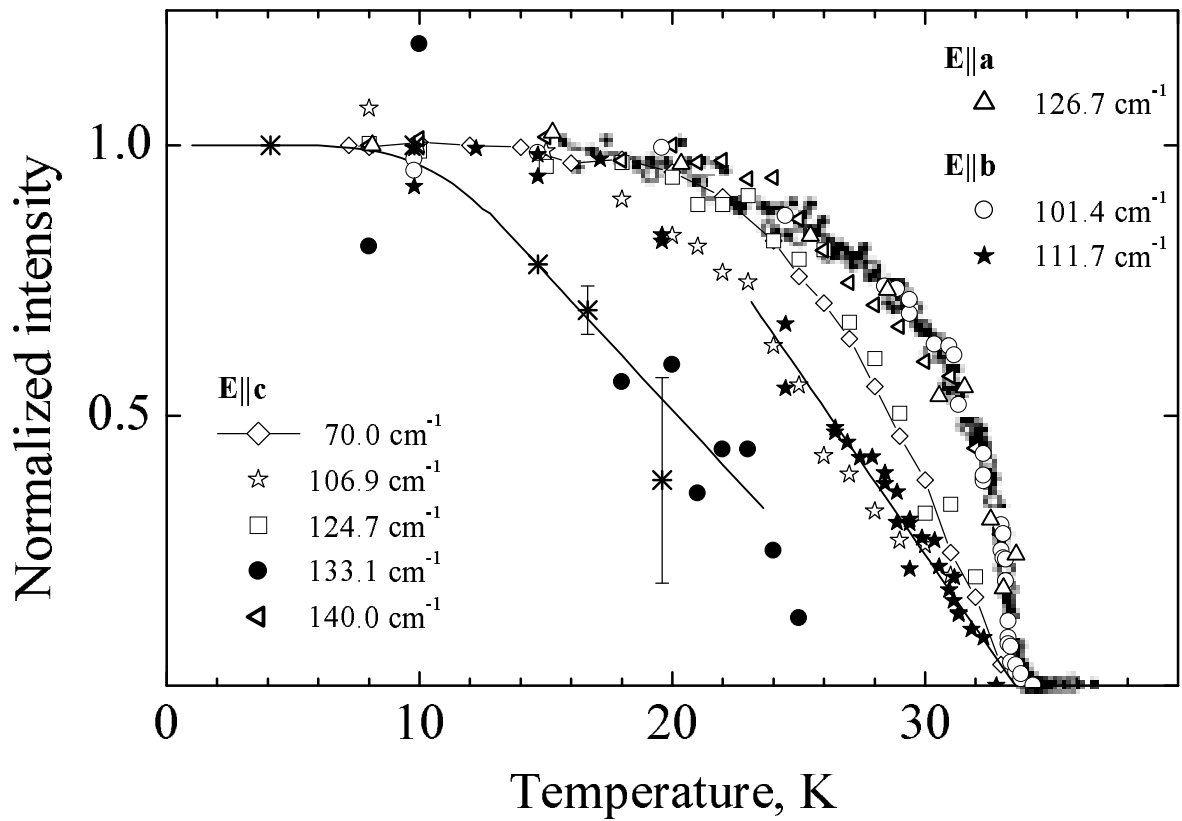
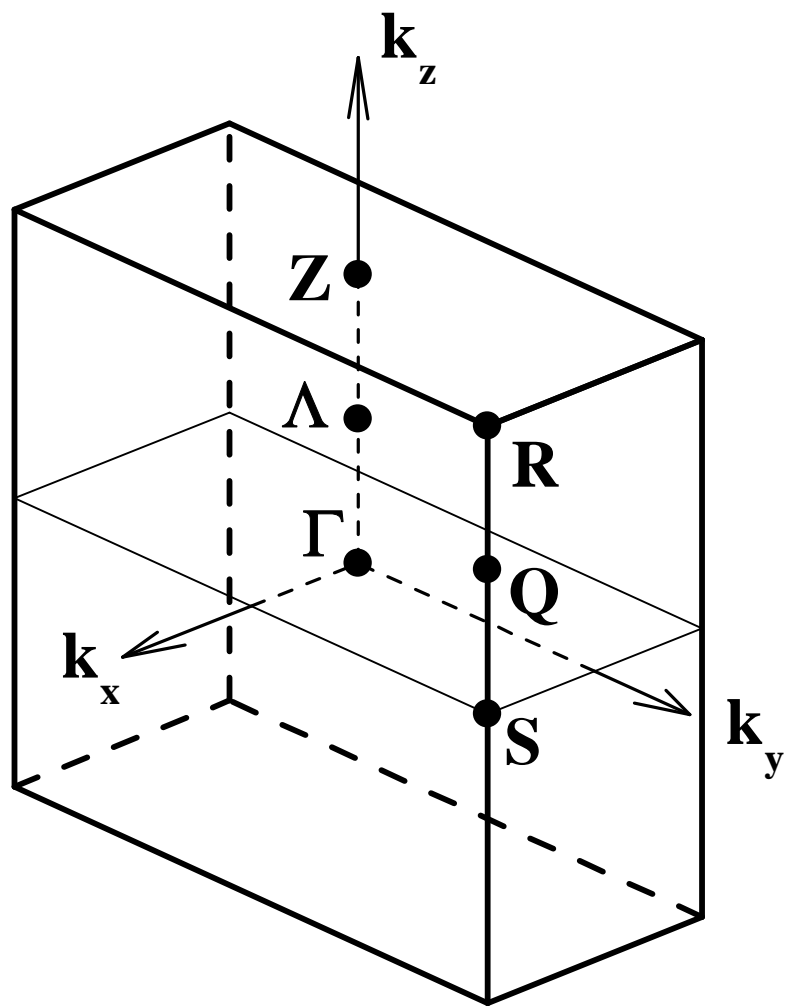
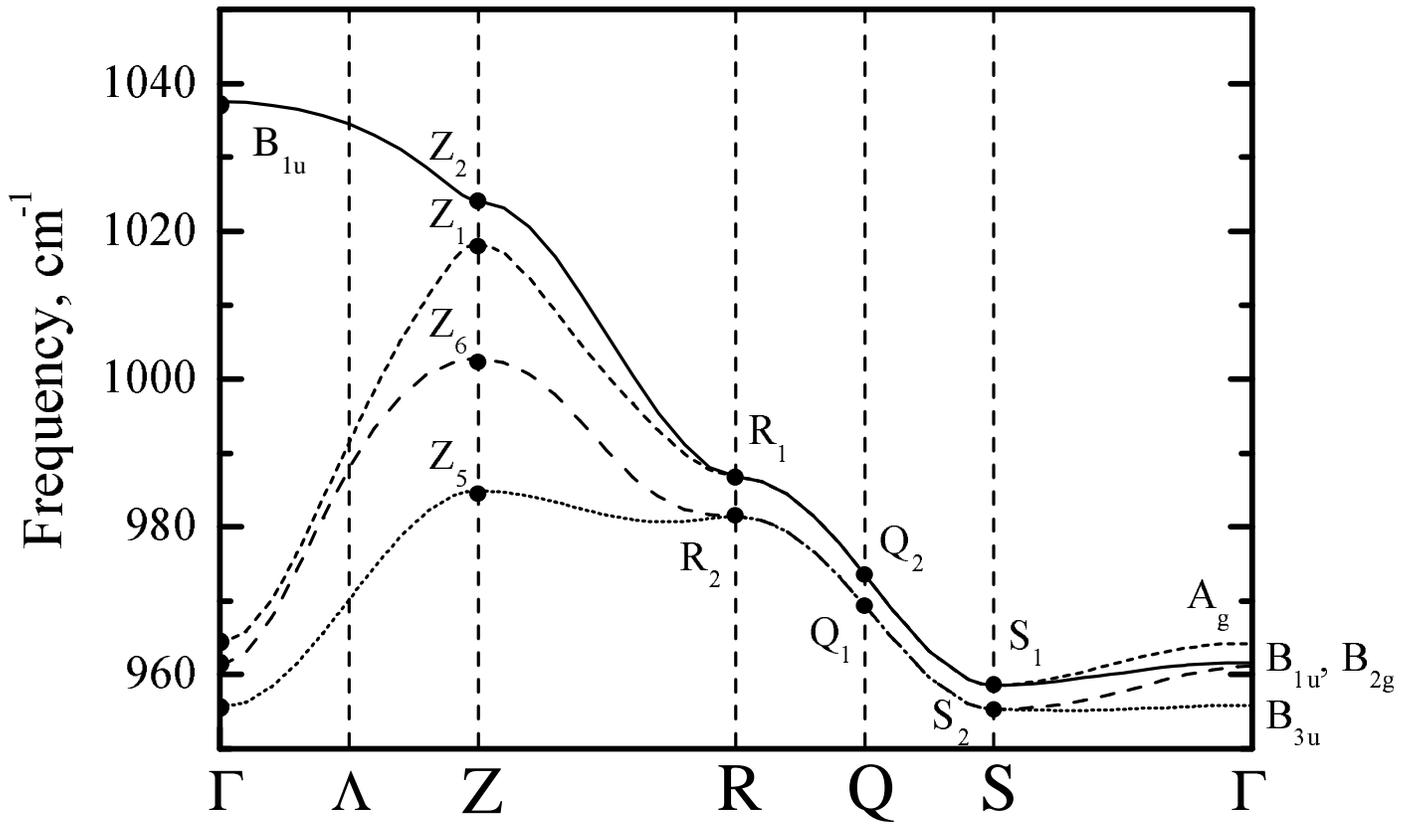
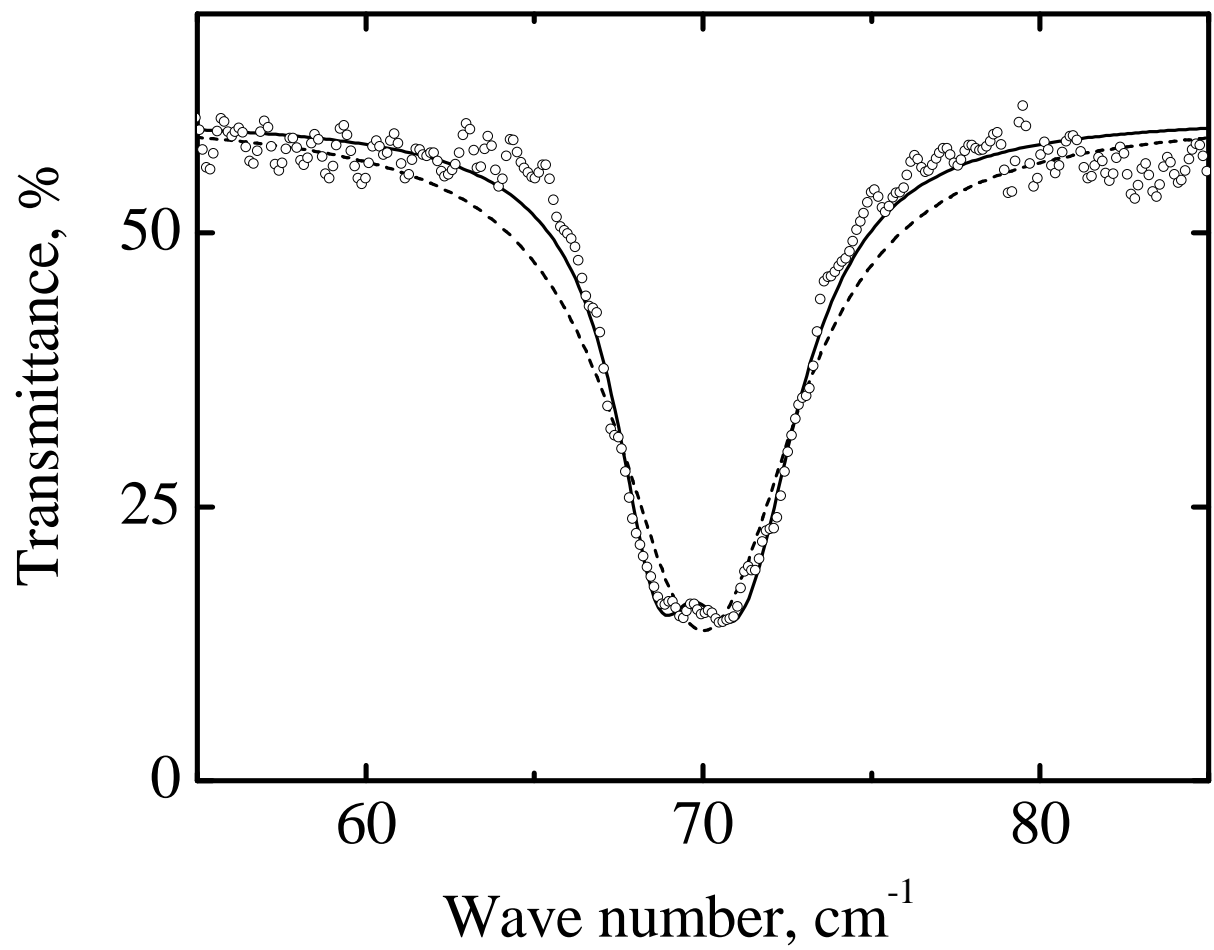
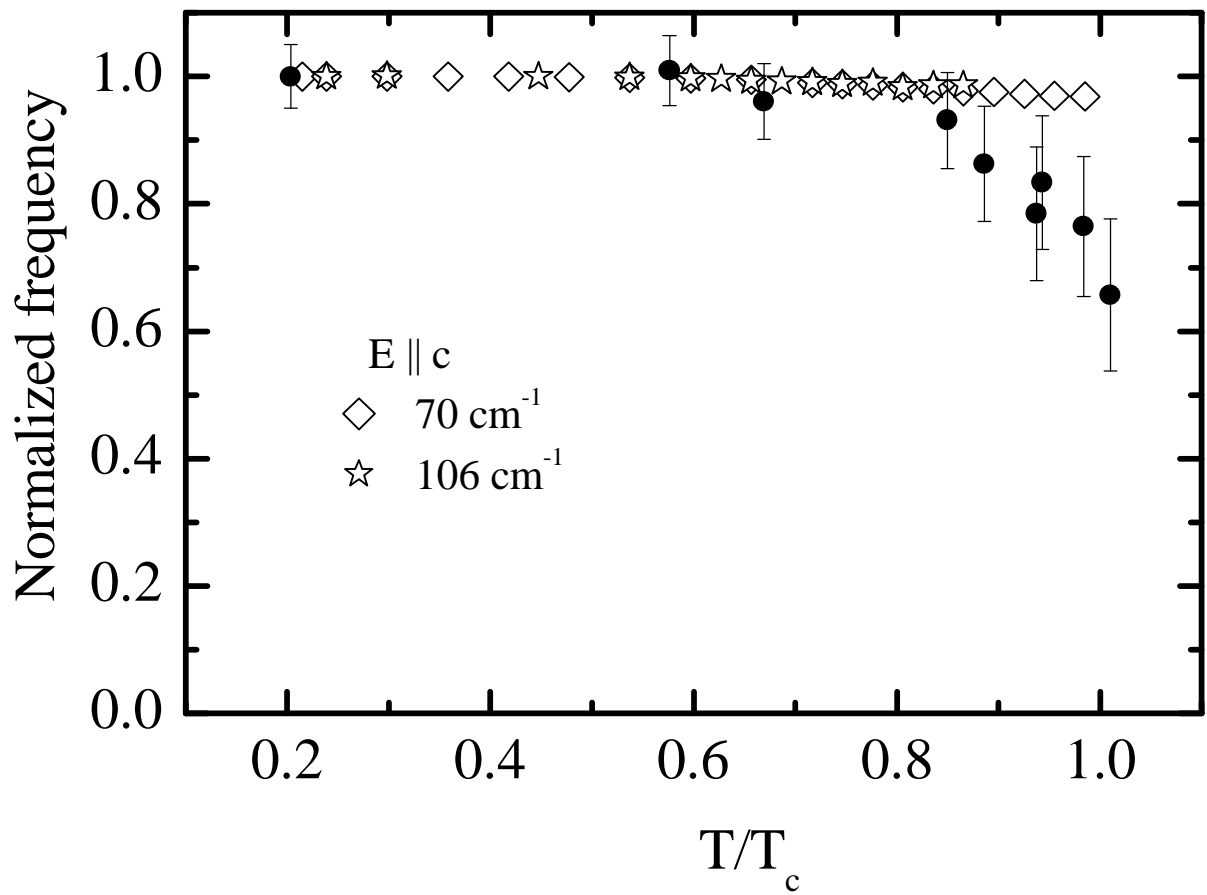
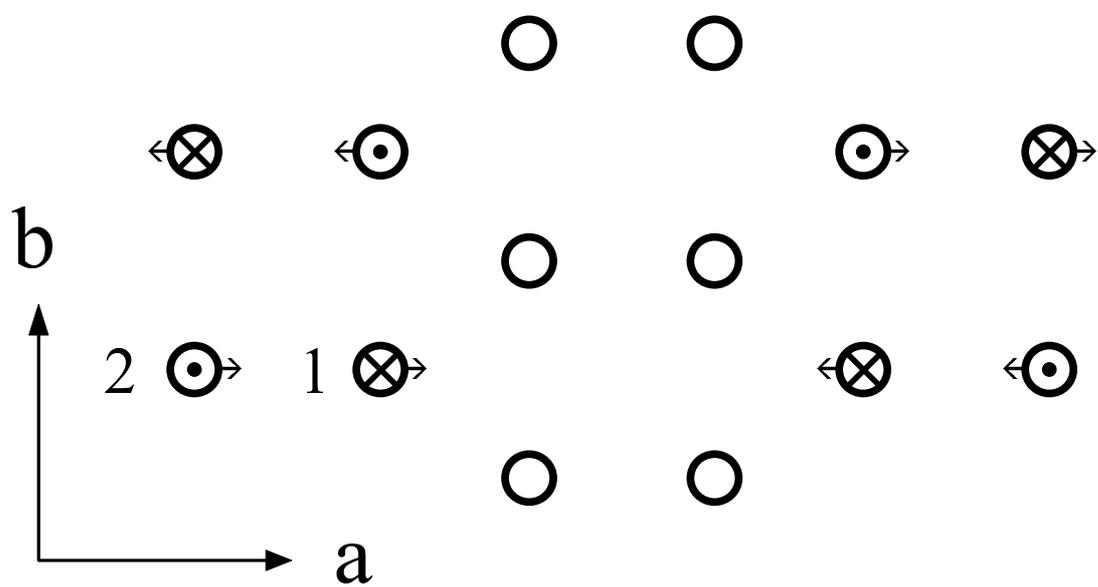
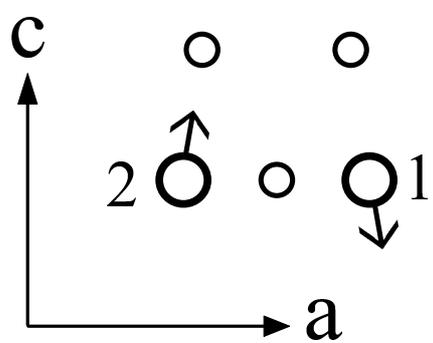
\end{document}